\documentclass[fleqn,10pt]{wlscirep}
\usepackage[utf8]{inputenc}
\usepackage[T1]{fontenc}
\usepackage{booktabs}
\usepackage{multirow}
\usepackage{comment}

\newcount\Comments 
\usepackage{color}
\usepackage{adjustbox}
        \usepackage{csquotes}
\definecolor{darkgreen}{rgb}{0,0.5,0}
\definecolor{purple}{rgb}{1,0,1}
\newcommand{\kibitz}[2]{\ifnum\Comments=1\textcolor{#1}{#2}\fi}

\newcommand{\petter}[1]{\kibitz{red}      {[PK: #1]}}

\newif\ifworkingversion

\title{On the impact of publicly available news and information transfer to financial markets}

\author[1]{Metod Jazbec}
\author[1]{Barna Pásztor}
\author[1]{Felix Faltings}
\author[2,3,*]{Nino Antulov-Fantulin}
\author[3,*]{Petter N. Kolm}

\affil[1]{Department of Computer Science, ETH Zurich, 8092 Zurich, Switzerland}
\affil[2]{Computational Social Science, ETH Zurich, 8092 Zurich, Switzerland}
\affil[3]{Courant Institute of Mathematical Sciences, New York University, New York 10012, USA}
\affil[*]{Corresponding authors: anino@ethz.ch, petter.kolm@nyu.edu}


\begin{abstract}
We quantify the propagation and absorption of large-scale publicly available news articles from the World Wide Web to financial markets. To extract publicly available information, we use the news archives from the Common Crawl, a nonprofit organization that crawls a large part of the web. We develop a processing pipeline to identify news articles associated with the constituent companies in the S\&P 500 index, an equity market index that measures the stock performance of U.S. companies. Using machine learning techniques, we extract sentiment scores from the Common Crawl News data and employ tools from information theory to quantify the information transfer from public news articles to the U.S. stock market. Furthermore, we analyze and quantify the economic significance of the news-based information with a simple sentiment-based portfolio trading strategy. Our findings provides support for that information in publicly available news on the World Wide Web has a statistically and economically significant impact on events in financial markets.
\end{abstract}

\begin{document}

\flushbottom
\maketitle
\thispagestyle{empty}

\section*{Introduction}
Studies of the impact of speculation and information arrival on the price dynamics of financial securities have a long history, going back to the early work of Bachelier\cite{bachelier2011louis} in 1900 and Mandelbrot\cite{Mandelbrot} in 1963 (see,  Jarrow and Protter\cite{jarrow2004short} for a historical account of these and related developments). In 1970, Fama\cite{Fama1970} formulated the efficient market hypothesis in financial economics, stating that security prices reflect all publicly available information.  Shortly after in 1973, Clark\cite{Clark1973} proposed the mixture of distribution hypothesis which asserts that the dynamics of price returns are governed by the information flow available to traders. Subsequently, novel models were introduced such as the sequential information arrival model\cite{SIAH}, news jump dynamics  \cite{jorion1988jump}, both in the 80s, and truncated Levy processes from econophysics\cite{stanley2000introduction} in the 90s, to name a few examples. With the rise of the World Wide Web and social media came an ever-increasing abundance of available data, allowing for more detailed studies of the impact of news on financial markets at different time-scales.\cite{chan2003stock, maheu2004news,vega2006stock, tetlock2007giving, gross2011machines, birz2011effect, vlastakis2012information, engelberg2012shorts, lillo2015news} Big data, coupled together with advancements in machine learning (ML)\cite{goodfellow2016deep} and complex systems research\cite{newman2011complex,Schreiber_2000,barnett2009granger,jizba2012renyi}, enabled more efficient analysis of financial data\cite{fang2016big}, such as web behaviour data\cite{lilloYahoo2016,heiberger2015collective,bordino2014stock,
piskorec2014cohesiveness, zhang2014internet,curme2015coupled,alanyali2013quantifying,dos2015breaking,yang2017genetic,ruiz2012semantic,wang2013financial,schumaker2009textual,amin2019sentiment,hisano_sornette2013}, social media\cite{ranco2015effects,rao2012analyzing,mao2011predicting,zheludev2014can,broadstock2019social}, web search queries\cite{bordino2012web,preis2013quantifying,zhang2013open}, online blogs\cite{de2008can, ruiz2012correlating} and other alternative data sources\cite{kolanovic2017big}. 

In this article, we use news articles from the Common Crawl News, a subset of 
the Common Crawl's petabytes of publicly available World Wide Web archives, to measure the impact of the arrival of new information about the constituent stocks in the S\&P 500 index at the time of publishing.  To the best of our knowledge, our study is the first one to use the Common Crawl in this way. We develop a cloud-based processing pipeline that identifies news articles in the Common Crawl News data that are related to the companies in the S\&P 500. As the Common Crawl public data archives are getting bigger, they are opening doors for many real-world ``data-hungry'' applications such as transformers models GPT\cite{brown2020language} and BERT\cite{wang2019construction}, a recent class of deep learning language models. We believe that public sources of news data is important not only for natural language processing (NLP) and finance communities but also for more general studies in complex systems and computational social sciences that are aiming to characterize (mis)information propagation and dynamics in techno-socio-economic systems. The abundance of high-frequency data around the financial systems enables complex systems researchers to have microscopic observables that allow verification of different models, theories, and hypotheses.  

With the use of ML methods from NLP\cite{blei2003latent,kelly2019TextData}, we analyze and extract sentiment from each news article in the Common Crawl repository, assigning a score in the range from zero to one that represent most negative (zero) through most positive (one) sentiment. To quantify the information propagation from the publicly available news articles on the World Wide Web to the companies in the S\&P 500, we use two different approaches. First, we employ tools from information theory of complex systems\cite{Shannon1948,Schreiber_2000,jizba2012renyi} to measure the impact of information transfer of the news sentiment scores on the returns of the constituent companies in the S\&P 500 index at an intraday level. Second, we implement and simulate the daily portfolio returns resulting from a simple trading strategy based on the extracted news sentiment score for each company. We use the returns from this strategy as an econometric instrument and compare it to several benchmark strategies that do not incorporate news sentiments. Our findings provides support for that information in publicly available news on the World Wide Web has a statistically and economically significant impact on events in financial markets.

\section*{Results}
\subsection*{The Common Crawl and financial news coverage }
The Common Crawl\footnote{\href{https://commoncrawl.org}{https://commoncrawl.org}} is a repository of web crawl data that is publicly available and accessible via Amazon Web Services. Since its start in 2011, the Common Crawl has continued to expand the number of websites it continuously accesses. Today, its crawl list contains more than three billion web pages.\footnote{Detailed statistics about the Common Crawl can found here: \href{https://commoncrawl.github.io/cc-crawl-statistics}{https://commoncrawl.github.io/cc-crawl-statistics}}
Recently, Common Crawl received a lot of attention for being the main data source for the training of OpenAI's GPT-3 language model, which has shown impressive results in the area of few-shot learning for natural language tasks \cite{brown2020language}.

Using the Common Crawl News, a subset of the Common Crawl exclusively for news articles,  we process and extract news articles related to the constituent companies in the S\&P500 index over the time period from 26th of August 2016 to 27th of February 2020. 
We choose the end of August 2016 as our starting point because prior periods have insufficient news coverage for the companies in the S\&P500. 
An article is matched to a company if and only if the company is mentioned in the title or the first paragraph (see the Supplementary Information (SI) for a detailed description of our data processing pipeline).

\begin{figure}[ht]
\centering
\includegraphics[width=\textwidth]{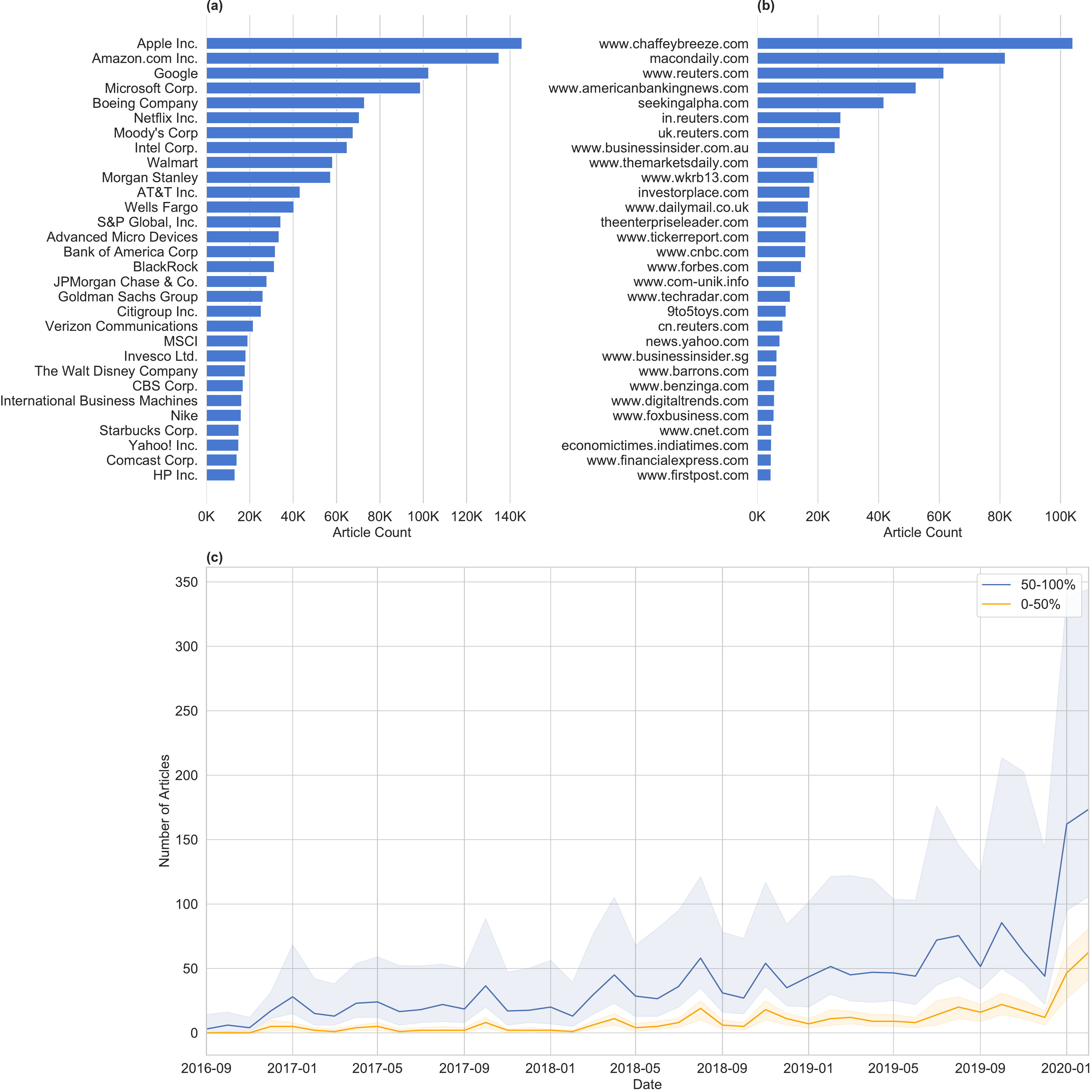}
\caption{Summary of the Common Crawl News dataset. (a) Most frequently mentioned companies as measured by the number of distinct articles. (b) Most frequent news sources as measured by the number of distinct articles associated with each source. (c) Median number of articles published per company and month. The companies are divided into top and bottom halves by the total number of articles published about them. The shaded regions represent the 25\% and 75\% percentiles of each half. 
}
\label{fig:exploratory_analysis}
\end{figure}
Figure~\ref{fig:exploratory_analysis}(a) shows the thirty most frequently occurring companies in our dataset as measured by the number of distinct articles that mention each company at least once. Figure~\ref{fig:exploratory_analysis}(b) shows the thirty top news sources as measured by the number of unique articles.\footnote{We omitted the domain \href{https://www.nbonews.com}{www.nbonews.com}. While the most frequently occurring source, it is no longer accessible and we were unable to verify its legitimacy. We provide summaries that include nbonews.com in the SI.} Not surprisingly, well-known publicly available financial news websites, such as \href{https://www.reuters.com}{www.reuters.com}, \href{https://www.seekingalpha.com}{www.seekingalpha.com}, \href{https://www.businessinsider.com}{www.businessinsider.com}, and \href{https://www.cnbc.com}{www.cnbc.com} appear amongst the most frequent sources. Other frequent sources, including \href{https://www.chaffeybreeze.com}{www.chaffeybreeze.com}, \href{https://www.macondaily.com}{www.macondaily.com}, and \href{https://www.americanbankingnews.com}{www.americanbankingnews.com} are less well-known. We note that  \href{https://www.chaffeybreeze.com}{www.chaffeybreeze.com} and \href{https://www.macondaily.com}{www.macondaily.com} redirect to \href{https://www.marketbeat.com}{www.marketbeat.com} that publish articles about specific companies and investor ratings. However, as Common Crawl only accesses publicly available sources with free content, any subscription based news services, such as the Wall Street Journal (\href{https://www.wsj.com}{https://www.wsj.com}) or Barron's (\href{https://www.barrons.com}{https://www.barrons.com}), are not part of our dataset. Figure~\ref{fig:exploratory_analysis}(c) depicts the median number of articles per company published each month throughout our time period. We divide the companies into upper and lower halves by total article count. The shaded regions show the 25\% and 75\% percentiles for each half by month, respectively. We observe that the distribution of published articles is right skewed as some companies have significant news coverage while others are mentioned infrequently. The shaded percentile regions illustrate that the top 50\% of the companies receive significantly more coverage than the bottom 50\%. In addition, we emphasize that the amount of news is increasing month over month, as the Common Crawl continues to increase the number of websites that are crawled. 

\subsection*{Extracting news sentiment from Common Crawl data}
Sentiment extraction refers to techniques from NLP that classify the polarity and sentiment of the expressed opinion in text-based documents as positive, negative or neutral. We deploy an extension of the SESTM (Sentiment Extraction via Screening and Topic Modeling)\cite{kelly2019TextData} model to assign sentiment scores to each article in our news dataset.   
After processing all articles, for each company $i$ in the S\&P500 we have an inhomogeneous time series of article sentiment scores occurring at irregular timestamps $\{t^i_1, t^i_2, \ldots ,  t^i_n\}$, that correspond to the publication times of $n$ articles. For our subsequent analysis we need the time series of sentiment scores to be occurring at regular time intervals. We achieve this by binning the sentiment series to hourly intervals and taking the average scores inside each bin. 
For example, for each company $i$ we obtain time series of hourly average sentiment scores  $\{s^i_t\}_{t=1}^{m}$, where $\{1, 2, \ldots, {m}\}$ represent the hourly-binned timestamps of total length $m$. 

\subsection*{Information transfer from news to stock returns}

To simplify the notation, we will drop the superscript index variable $i$, that are corresponding to variables $s^i_t$ for a particular stock or company $i$. Hence the hourly price returns of a particular stock will be denoted by $r_t$.  
We characterize the information transfer from the news sentiment series of each company in the S\&P500 index to its stock price returns by measuring its \emph{transfer entropy} (TE) \cite{Schreiber_2000,jizba2012renyi}, an entropy-based measure from information theory\cite{Shannon1948}. 
In particular, we use transfer entropy to quantify the amount of uncertainty reduction in the future return, $r_{t+1}$, for each stock given the information of lagged news sentiment and price returns, $(s_t,r_t)$.  
From an information theoretic perspective, TE is the difference in the number of bits needed to encode the information of the state of the process by the transition probabilities $p(r_{t+1}| r_t)$ instead of $p(r_{t+1}| r_t, s_t)$. 
To address any non-stationarity of the sentiment processes, we perform an augmented Dickey–Fuller test and compute first differences when necessary. To compute the statistical significance (p-values) of TE, we employ a non-parametric bootstrap method of the underlying Markov process \cite{TEpval,TE_R} and use effective TE\cite{ETE} to perform finite sample corrections.

\begin{figure}
\centering
\includegraphics[width=\textwidth]{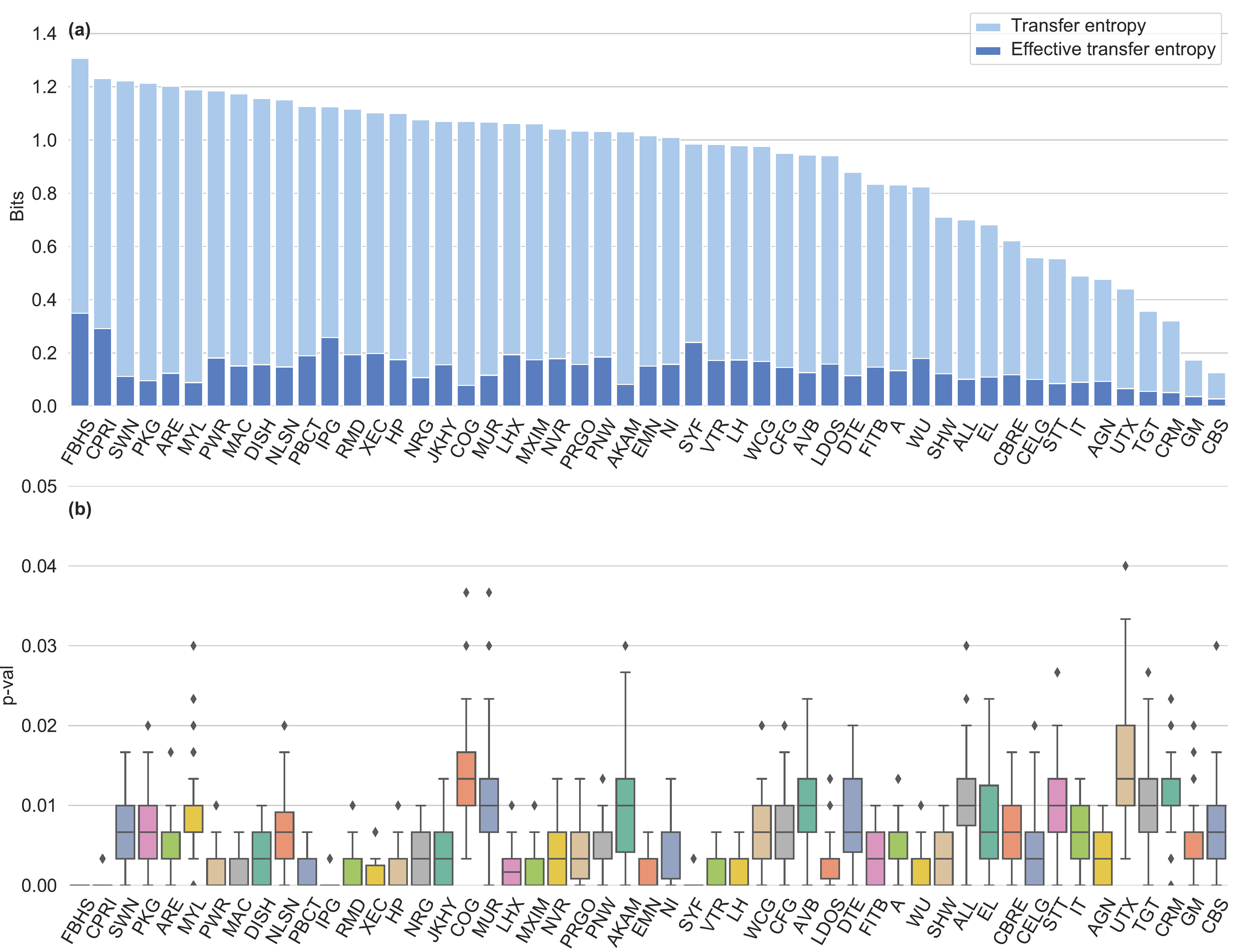}
\caption{{\bfseries Panel (a)} shows companies and corresponding significant Shannon transfer entropy (and effective transfer entropy) from hourly sentiment score differences to hourly price returns. The unit of transfer entropy is bits (logarithm with base 2), corresponding to the reduction of the average optimal code length needed to encode stock returns with lagged sentiment.  Transfer entropy was calculated for the period from January 2018 through February 2020 using time series of hourly returns from 9:30 a.m. to 15:30 p.m. Eastern Time and corresponding lagged average sentiment scores. The statistical significance (p-value$<0.01$) of transfer entropy was estimated with 300 bootstrap samples and 100 shuffles to obtain the effective transfer entropy. {\bfseries Panel (b)} depicts box and whisker plots of estimated distributions of the p-values for selected company tickers. The box and whisker plots show Q1, median, Q3, minimum, maximum and estimated outliers. 
 }
\label{fig:transfer_entropy}
\end{figure}

\begin{figure}
\centering
\includegraphics[width=\textwidth]{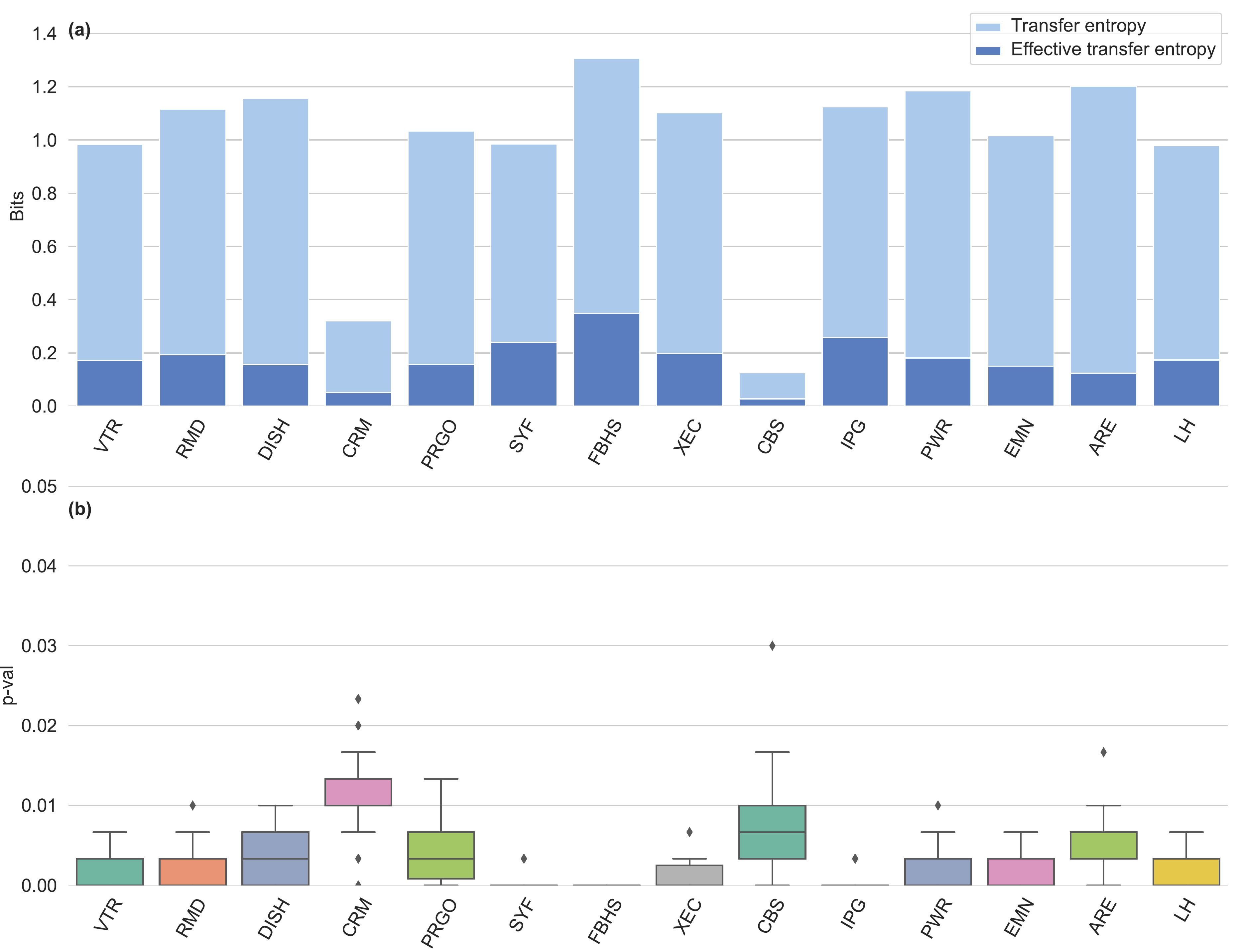}
\caption{{\bfseries Panel (a)} shows companies and corresponding Shannon transfer entropy and effective transfer entropy after false discovery rate control (FDR$<0.05$) using the Benjamini–Yekutieli procedure. The unit of transfer entropy is bits (logarithm with base 2), corresponding to the reduction of the average optimal code length needed to encode stock returns with lagged sentiment. Transfer entropy was calculated for the period from January 2018 through February 2020 using time series of hourly returns from 9:30 a.m. to 15:30 p.m. Eastern Time and corresponding average sentiment scores. {\bfseries Panel (b)} depicts box and whisker plots of estimated distributions of the p-values for selected company tickers. The box and whisker plots show Q1, median, Q3, minimum, maximum and estimated outliers.}
\label{fig:transfer_entropy_FDR}
\end{figure}

\begin{figure}
\centering
\includegraphics[width=\textwidth]{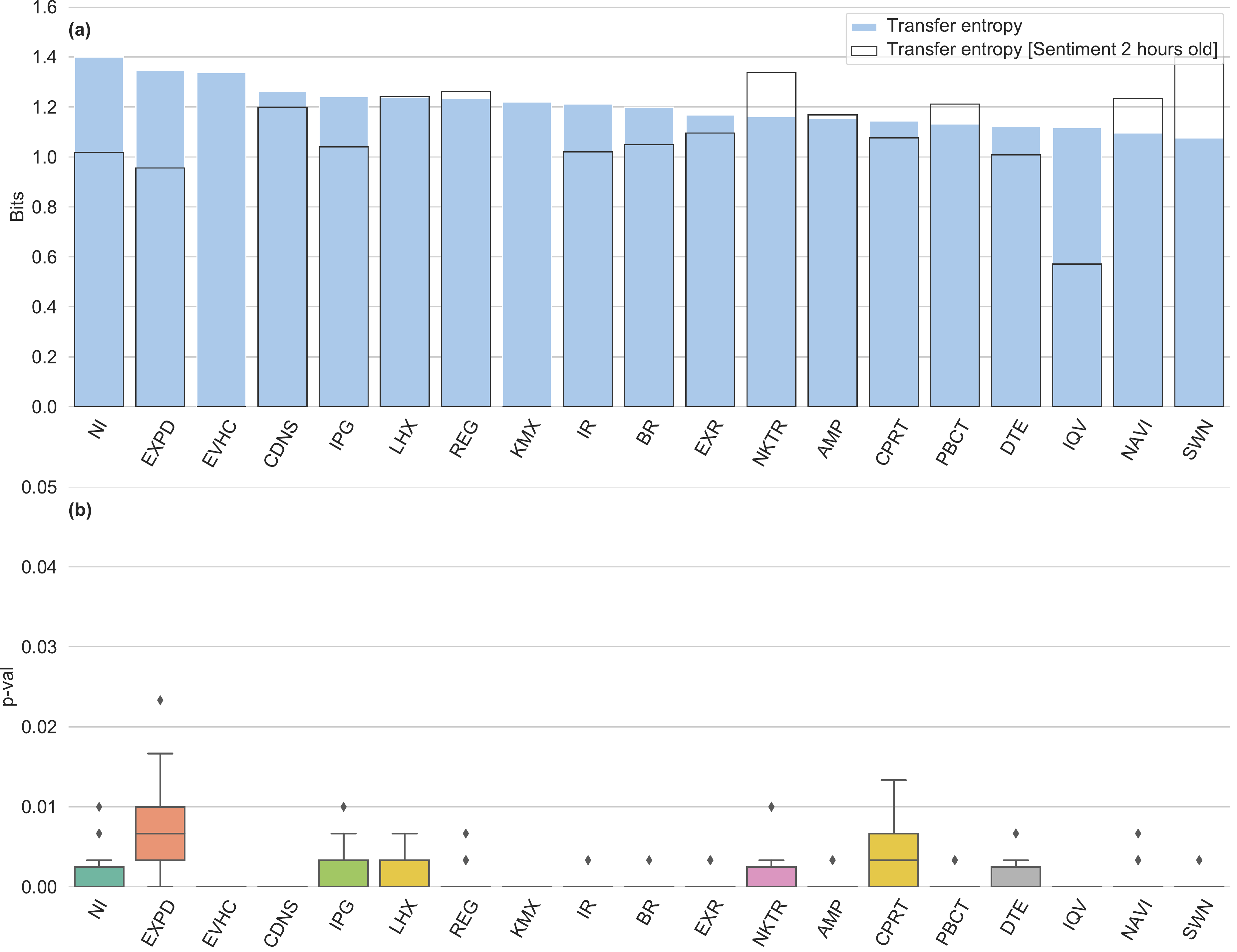}
\caption{\textbf{Panel (a)} shows of companies and corresponding Shannon transfer entropy and effective transfer entropy (Jan 2018 until Feb 2020) with two hours old sentiment with the false discover rate control (FDR$<0.05$) by using Benjamini–Yekutieli procedure. The unit of transfer entropy is bits (logarithm with base 2), corresponding to the reduction of the average optimal code length needed to encode stock returns with lagged sentiment. \textbf{Panel (b)} depicts box and whisker plots of estimated distributions of the p-values for selected company tickers. The box and whisker plots show Q1, median, Q3, minimum, maximum and estimated outliers.
}
\label{fig:transfer_entropy_lag_BY}
\end{figure}

Using our dataset over the period from 3rd of January 2018 until 27th of February 2020\footnote{A period between August 2016 and December 2017 is used as a ``warm-up'' period for the sentiment model. For the exact fitting procedure see the SI.}, for each company we compute TE between hourly sentiment score differences and hourly price returns. Figure~\ref{fig:transfer_entropy}(a) and (b) display companies with statistically significant TE (p-val$<0.01$) and the estimated distribution of p-values. We observe that the distribution of  p-values of the depicted stocks are below the 0.05 level.  To control the false discovery rate (FDR), we apply the Benjamini–Yekutieli procedure\cite{benjaminiYekutieli} (FDR$<0.05$). Figure~\ref{fig:transfer_entropy_FDR} shows the results after the Benjamini–Yekutieli procedure. These results suggest the presence of statistically significant reduction of uncertainty for the considered time period in stock returns when using historical hourly sentiment from public news.
Next, we analyze whether older sentiment signals are carrying valuable information content for the reduction of price returns uncertainty. In particular, in Figure \ref{fig:transfer_entropy_lag_BY} we show the set of companies that have significant TE with two hours old sentiment and Benjamini–Yekutieli correction procedure\cite{benjaminiYekutieli} (FDR<0.05). Again, we find the presence of statistically significant reduction of uncertainty when two hour old sentiment signals are being used. Our analysis corroborates public news contribution to information dissemination and price discovery at different time-scales\cite{muller1997volatilities}. The existence of multiple time-scales in financial markets is connected to a heterogeneous market hypothesis\cite{muller1997volatilities} and arises from differences in heterogeneity across traders, including different trading constraints, risk profiles, locations, and information processing and decision frequencies.\cite{corsi2009simple}

\subsection*{Economical significance of news sentiment from Common Crawl data}
In the previous section we demonstrated that public news sentiment has a statistically significant impact on the uncertainty reduction of future stock returns. This result poses the question whether this impact is also economically significant. To address this question, we analyze performance of the following simple sentiment-based trading strategy. For every day we rank all S\&P 500 companies based on their news sentiment scores from news articles published between 9:30 a.m. on the previous trading day and 9:00 a.m. of the current day, where times reflect Eastern Time (ET) throughout this article. For companies with multiple news articles, we average their news sentiments to obtain a single sentiment score for each company. Daily, we form a portfolio with long positions of equal amounts in the 20 companies with the most positive sentiment scores and short positions of equal amounts in the 20 companies with the most negative sentiment scores.\footnote{A long position refers to the investor having bought and therefore owns shares. A short position refers to the investor having borrowed and sold shares on the open market, planning to buy them back later for less money.}  We refer to this daily rebalanced portfolio as the \emph{Day 1 sentiment strategy}, where ``Day 1'' denotes the one-day lag of the sentiment scores. We track its daily open-to-open return over time, i.e. from 9:30 a.m. on the day of portfolio formation to 9:30 a.m. the following day. Similarly, we form Day 0 and  Day -1 sentiment portfolios and track their daily returns through time.\footnote{Specifically, denoting the current trading day by $t$, we use news from 9:30 a.m. on day ${t-1}$ through 9:00 a.m. on day $t$ for the Day 1 portfolio, 9:30 a.m. on day ${t}$ through 9:00 a.m. on day ${t+1}$ for the Day 0 portfolio and 9:30 a.m. on day ${t+1}$ through 9:00 a.m. on day ${t+2}$ for the Day -1 portfolio. For all portfolios we enter the market at 9:30 a.m. on day $t$ and hold our positions until 9:30 a.m. on day ${t+1}$.} Note that the Day 0 and Day -1 portfolios are not tradable in practice as they rely on receiving news ahead of its publishing time. However, we use these ``look-ahead'' portfolios for comparison purposes below. 

We compare our sentiment-based trading strategies performance to the SPDR S\&P 500 trust (ticker symbol: SPY) and a set of random portfolios as a null-model benchmark. SPY is an exchange-traded fund tracking the S\&P 500 index. Each random portfolio is rebalanced daily at the same time as the Day 1 sentiment strategy and consists of long and short legs, each with positions of equal amounts in 20 randomly chosen stocks from the S\&P 500.  We simulate 500 random portfolio histories and use their resulting return series to bootstrap performance metrics.  

\begin{table}[ht]
\setlength{\tabcolsep}{7pt}
\centering
  \begin{tabular}{llllll}
    \toprule
    \multirow{5}{*}{} &
      \multicolumn{3}{c}{\textbf{Common Crawl}}  &
      \multicolumn{1}{c}{} &
      \multicolumn{1}{c}{}\\
      & {Day -1} & {Day 0 } & {Day 1} & {\textbf{SPY}} & {\textbf{Random}}  \\
      \midrule
    Ann. avg. return & 48.95\%  & 45.42\% & 21.02\%  & 7.25\% & $-0.11 \pm 5.67\% $  \\[0.1cm]
    Ann. volatility &  11.65\%  &  12.43\% &  12.85\%  &  15.06\% & $8.37 \pm 0.33\%$  \\[0.1cm]
    Ann. Sharpe ratio (p-value) & 4.20 (< 0.001) & 3.66 (< 0.001) & 1.64 (< 0.01)  & 0.48 & $ -0.01 \pm 0.68$ \\[0.1cm]
    MDD & 8.82\%  & 8.34\% & 10.31\%  & 21.04\% & $14.64 \pm 6.19\% $  \\[0.1cm]
    
    Ann. $\alpha$ (p-value) & 47.88\% (< 0.001)  & 45.36\% (< 0.001) & 20.69\% (0.02)  & 0  & $ 0.05 \pm 5.70\% $\\[0.1cm]
    $R^2$ & 0.004  & 0.001 & 0.004  & 1 & $ 0.002 \pm 0.004 $ \\
    \bottomrule
  \end{tabular}
  \caption{\label{tab:trading_results} Performance statistics of the Day 1 sentiment trading strategy and benchmarks from January 2018 through February 2020. The sentiment trading strategy is based on news articles from the Common Crawl News dataset. SPY is the SPDR S\&P 500 trust. ``Random'' denotes the baseline strategy where each day we randomly select companies to invest in. ``Day 0'' and ``Day -1'' are ``look-ahead'' sentiment strategies, reported for comparison purposes. Statistics are computed using daily returns (n=542). MDD is the Maximum Daily Drawdown defined as the maximum observed decline from a historical peak of the price until a new peak is attained. The p-values for the Sharpe ratios were bootstrapped from 500 random backtests. We obtain $\alpha$ (the intercept) and $R^2$ by regressing the daily returns of the portfolios on the daily returns of the SPY. The performance metrics of the random portfolios were bootstrapped from 500 random backtests.}
\end{table}

\begin{figure}[ht]
\centering
\includegraphics[width=1.0\linewidth]{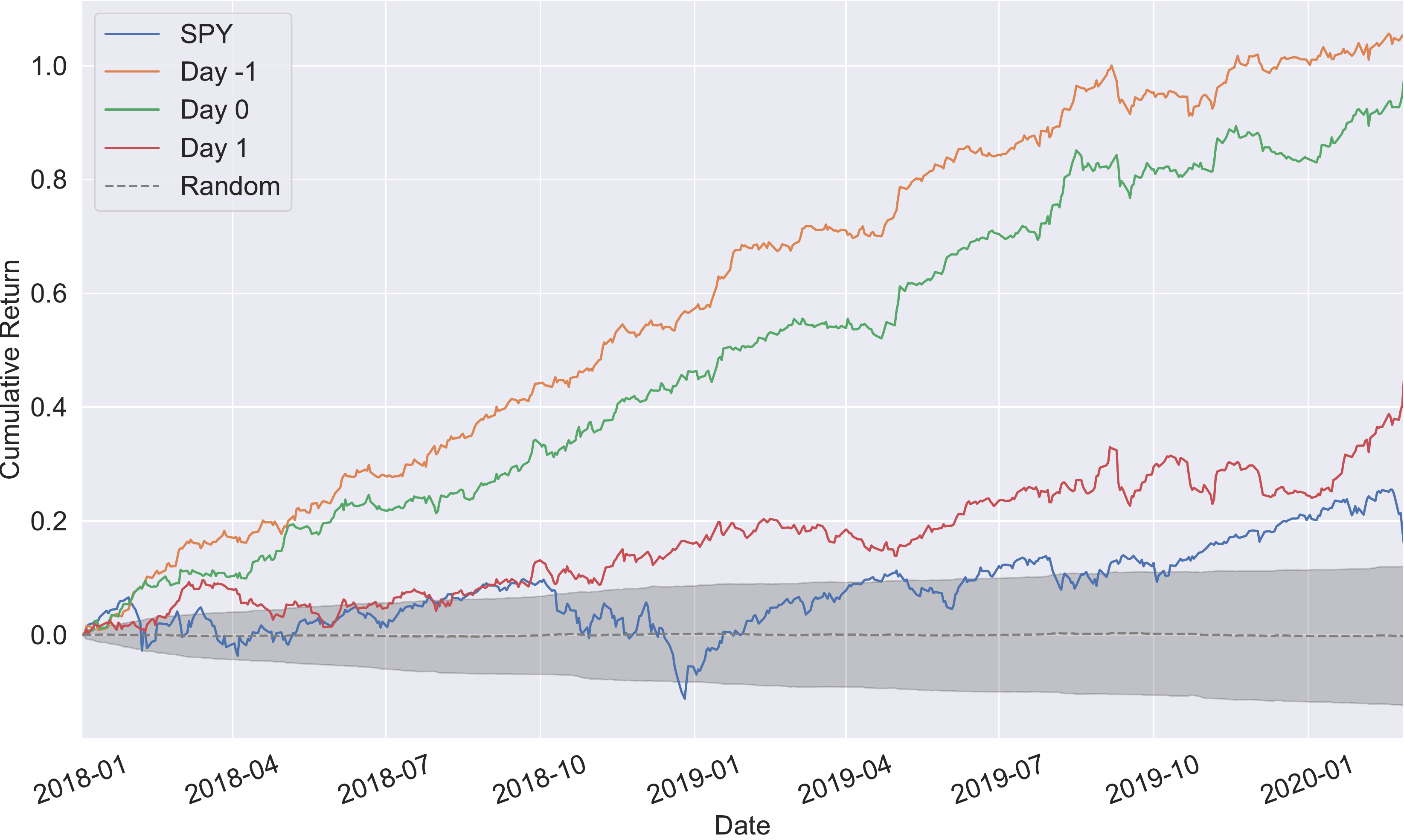}
\caption{
Cumulative returns of trading strategies and benchmarks. ``Day 1'' represents the cumulative returns of the Day 1 sentiment strategy based on the Common Crawl dataset from January 2018 through February 2020. SPY is the SPDR S\&P 500 trust. ``Random'' denotes the average of the random strategies along with one standard deviation confidence bands obtained from 500 simulations. ``Day 0'' and ``Day -1'' are the ``look-ahead'' sentiment strategies relying on future information.} 
\label{fig:trading_results}
\end{figure}

In Figure~\ref{fig:trading_results} we depict the cumulative return of the Day 1 sentiment strategy relative to our benchmarks from January 2018 through February 2020. We summarize the performance statistics of the Day 1 sentiment trading strategy and benchmarks in Table~\ref{tab:trading_results}. 
We use the annualized Sharpe ratio as one of performance metric, defined as the annualized average return divided by the annualized volatility of return. Clearly, the Day 1 sentiment strategy outperforms the SPY and random portfolios, obtaining an annualized Sharpe ratio of 1.64 as compared to 0.48 and -0.01 for SPY and random portfolios, respectively. Regressing the returns of the Day 1 sentiment strategy on the returns of SPY, we observe that the intercept, denoted by $\alpha$, and $R^2$ of this regression is 20.69\% (annualized) and 0.4\%, respectively. $\alpha$ is significant at the 1\%-level. We conclude that that the Day 1 sentiment strategy (a) outperforms the market and (b) is uncorrelated with the market. This supports that there is economically and statistically significant information in public news sources. 
However, note that contrary to the transfer entropy measure that quantifies the average uncertainty reduction during the whole period, the trading strategy is an econometric method that is based on a dynamically re-balanced portfolio that adapts to the arrival of new information through time. 

As expected, we note that both the Day -1 and Day 0 strategies outperform the Day 1 strategy, achieving Sharpe ratios of 4.20 and 3.66, respectively. Similarly, they strongly outperform both baselines and display no significant correlation with the market ($R^2$ when regressing on SPY returns is 0.4\% and 0.1\%, respectively). While these two strategies cannot be traded in practice as they rely on future information, they illustrate further that there is significant correlation between stock returns and sentiment derived from the Common Crawl News dataset. 

We emphasize that there are many ways in which information about news sentiment can be traded in financial markets. Of course, the simple trading strategy we chose here is not representative of those used by sophisticated investors such as professional money managers and hedge funds. Nevertheless, while we performed our analysis using this basic trading strategy, we expect that more sophisticated approaches would only strengthen our findings. Additionally, in real-world trading it is important to consider trading costs and more precise risk-adjusted return contributions, but such an analysis is beyond the scope of this article. 

\subsection*{Comparing the information transfer of private and public news to financial markets}
We investigate to what extent there is a difference in the information transfer of publicly available news as compared to commercially available news data to financial security prices. As noted above, the Common Crawl data consists of only freely available public news. That is, the dataset does not contain news content from web pages and news providers that are subscription-based or require registration.

To measure the effect of non-public news sources, we obtained the Alexandria dataset\footnote{\href{https://www.alexability.com}{https://www.alexability.com}}, a commercial news database consisting of financial news curated from about a dozen subscription based data sources,  including Dow Jones News Wire, Wall Street Journal and Barrons. Alexandria uses a proprietary algorithm to associate each news article to the companies in question and to assign sentiment scores.  

We use the same simple trading strategy as above to build daily sentiment portfolio using the Alexandria dataset and compare its trading results to those based on the Common Crawl dataset. The trading strategy using the Alexandria dataset obtains a Sharpe ratio of 1.51 over our time horizon (other performance metrics are available in the SI). Importantly, the correlation of the return series of the two strategies is only 0.07 (p-val<0.1). 
That the correlation is not statistically different from zero suggests the sentiment scores derived from the Common Crawl and Alexandria datasets are based on different underlying information. In fact, the average Jaccard index\footnote{The Jaccard index, a measure of the similarity between two discrete sets $A$ and $B$, is defined as $J(A, B) := |A \bigcap B| / |A \bigcup B|$. It takes values in the range $[0,1]$. The higher the index, the greater the similarity between the two sets.}, between the long and short stock positions based on the sentiment scores from the Alexandria and Common Crawl datasets are $0.020 \pm 0.022$ and $0.019 \pm 0.023$, respectively. This means that the overlap of companies in the portfolios formed on the Alexandria and Common Crawl sentiment scores is less than one on average.

We conclude there is valuable information present in both datasets to predict future returns. However, the information from the datasets are different and the Common Crawl News dataset provides complementary information to that of the Alexandria dataset. There are two main reasons for why the datasets are different: (a) they use different news sources, and (b) the computed sentiment scores are determined from different models. Alexandria relies on subsctiption-based financial news, whereas the Common Crawl only accesses publicly available sources. Alexandria deploys a proprietary ML approach to compute sentiment scores for each news article, whereas we use the SESTM model to determine sentiment scores for the news articles from the Common Crawl.

\section*{Discussion}
Processing roughly 400 million articles from the Common Craw News data comes with many non-trivial engineering challenges, including parsing different HTML formats used on the websites, identifying and removing duplicate articles, aligning each and financial alignment to corresponding companies. 
The SESTM sentiment model was chosen as a consensus of the complexity, interpretability, and theoretical foundations in supervised learning and topic modeling in NLP. In the SI, we explore the effects of deep learning models for sentiment extraction, using the pre-trained\cite{araci2019finbert} Bidirectional Encoder Representations from Transformers model (BERT)\cite{devlin2018bert}. 
It is important to emphasize that the focus of this paper is not on what the best sentiment model is, but rather on the analysis of the interaction of news from the World Wide Web and the financial market as a prototype of an efficient ``information absorbing'' system. By analyzing the time-series of sentiment scores and price returns, we find evidence of statistically significant transfer of information on the intraday level over the period from January 2018 through February 2020. 
In this study, we did not analyze possible cofounding effects~\cite{PhysRevE.72.026222, vakorin2009confounding} between multivariate time series of sentiment and stock price returns, as we were not focused on causal inference~\cite{sugihara2012detecting,yang2018causal } but on bivariate transfer of information between news to corresponding stocks. 
The multivariate case of transfer entropy can be further extended with partial transfer entropy~\cite{kugiumtzis2013partial,PhysRevE.72.026222} in future work. By using a simple sentiment-based trading strategy as an econometric tool, we find that our sentiment signals from public news are carrying both economic value and complementary information compared to non-public and commercial news data providers. Our analysis provide support for that public news contribute to information dissemination and price discovery at different time-scales during a single trading day.  
Finally, along with this study, we are releasing one of the largest corpora of processed financial documents for wide research use (see the SI for information on how to obtain a copy of the dataset). 

\section*{Methods}

\subsection*{Common Crawl data}
\begin{figure}[ht]
\centering
\includegraphics[scale = 0.75]{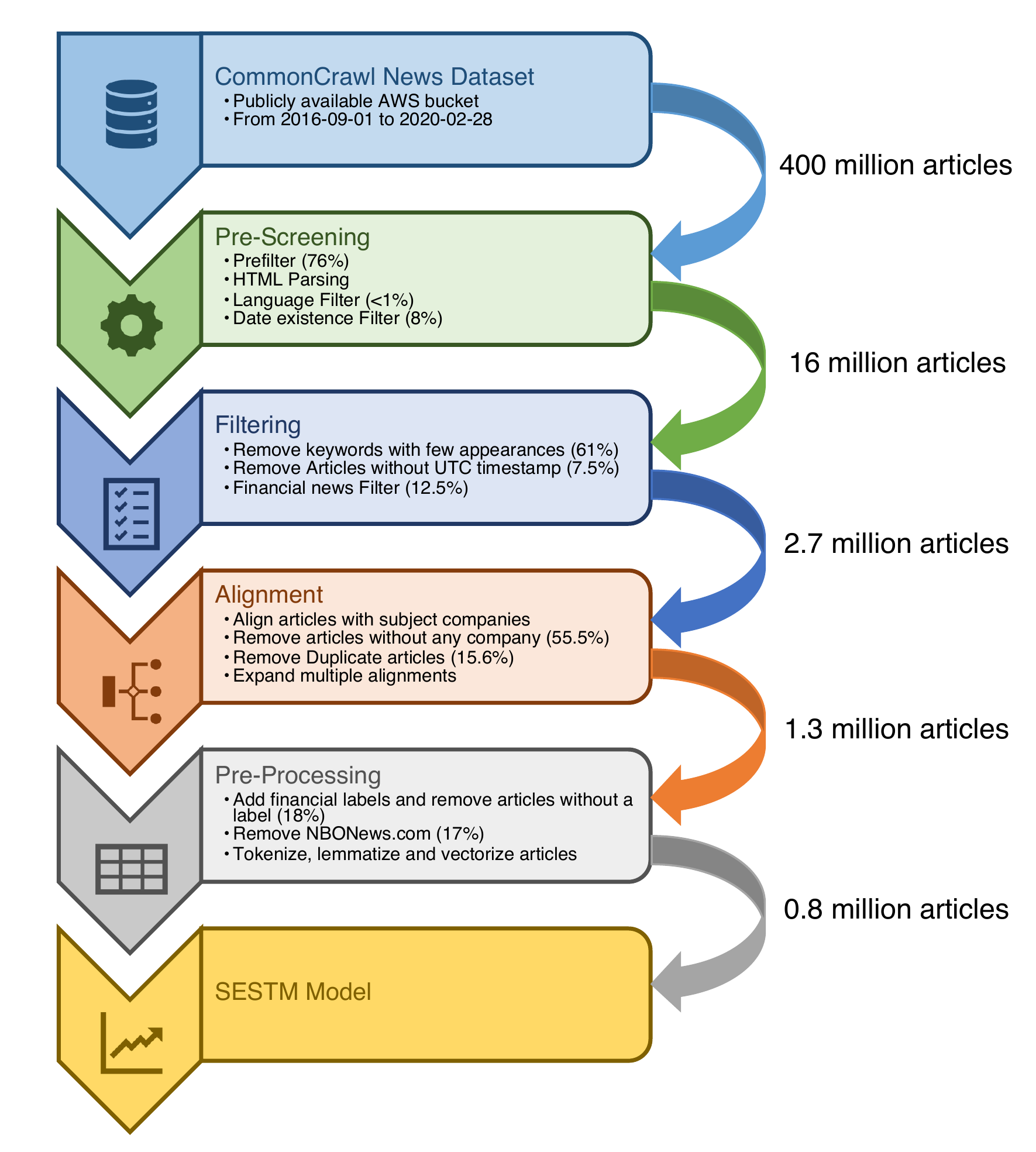}
\caption{The pipeline deployed to process and transform the Common Crawl News dataset into the dataset used by the sentiment model. Each box represents one stage of the pipeline where data transformation and filtering steps are applied. The numbers next to the arrows show how many articles are passed on from one stage to the next. The percentages in the brackets after each filtering step show the proportion of articles removed in that specific step. 
}
\label{fig:process_pipeline}
\end{figure}

News data is freely and publicly available from many online sources including newspapers, news outlets, and news aggregators. Assembling a dataset that covers a representative subset of these sources is a major task. For example, it took about three and a half years for Google to develop a release ready version of Google News.\footnote{See, \href{https://googleblog.blogspot.com/2006/01/and-now-news.html}{https://googleblog.blogspot.com/2006/01/and-now-news.html}}  
In this article we use a datatset from the Common Crawl, a nonprofit organization that collects data from the World Wide Web and provides it for free to the public.

Since its start in 2008, the Common Crawl  has collected petabytes of data and their developers have continued to improve their system and expand the number of websites visited by their crawler. We performed our analysis using the Common Crawl News, a subset of the Common Crawl dataset containing only news articles.\footnote{This version of the dataset is available at \href{https://commoncrawl.org/2016/10/news-dataset-available}{https://commoncrawl.org/2016/10/news-dataset-available}} From its start in 2016 until the end of February 2020, Common Crawl collected about eight terabytes of news from publicly available news sources on the World Wide Web, corresponding to a total of about 400 million articles covering a wide range of topics in different languages. The coverage of the Common Crawl News dataset has continued to expand over time as new sources are added to the crawler.

Figure~\ref{fig:process_pipeline} depicts the steps we took in processing the Common Crawl News dataset. Processing the Common Crawl News dataset is complicated due to its size, the content's raw format, the different HTML formats used on websites and the broad coverage of topics and languages. Furthermore, as part of the processing we had to address several issues including (a) extracting the main text and additional meta-information such as timestamps etc.~from HTML strings of each article, (b) matching each article to the relevant company names and tickers, (c) decide which articles relate to the financial performance of the subject company, and (d) removing any duplicate news articles present in the data.  
After the processing we have a dataset of about 1.3 million financial articles written in English covering S\&P 500 companies from August 2016 through February 2020. For each article, we identify its URL, publication time, crawling time, title and all companies referenced. In addition, for each article, we determine the company for which each news article  is most pertinent. We describe the details of our processing pipeline and the specific schema of the dataset in the SI.

\subsection*{Transfer entropy}
To avoid making specific assumptions of the relationship between sentiment and stock returns,  rather than using classical Granger causality we use transfer entropy, a model-free measure  from information theory that is not restricted to linear dynamics or Gaussian assumptions.\cite{TEpval,barnett2009granger} For a random variable $X$, the Shannon entropy $H(X)=\mathbb{E}[-\log p(X)]$ measures the expected level of ``information'' or ``uncertainty'' associated with its outcomes. Intuitively, if the logarithm is expressed in base two then $H(X)$ represents the number of bits of the optimal code length for a lossless data compression of events from data source $X$. In order to quantify information content $H( X_t )$ from a time dependent stochastic process, $\{ X_t \}$, one needs to analyze transition probabilities\cite{HLAVACKOVASCHINDLER2007} of the underlying stochastic process. In particular, using the idea of finite-order Markov processes, Schreiber\cite{Schreiber_2000} introduced the transfer entropy (TE) measure, which detects information transfer between systems evolving through time.  
For each company in the S\&P 500 we compute the \emph{transfer entropy} (TE) \cite{Schreiber_2000,jizba2012renyi} from the company news sentiment $\{ s_t \}$ to stock returns $\{ r_t \}$, defined as
\begin{equation}\label{eq:transfer-entropy}
   TE_{s \to r} := H(r_{t+1}| r_t ) - H(r_{t+1}| r_t, s_t), 
\end{equation}
where $H(X|Y):=-\sum_{i,j} p(x_i,y_j) \log[p(x_i|y_j)]$ 
denotes the conditional Shannon entropy.\cite{jizba2012renyi} The transfer entropy \eqref{eq:transfer-entropy} can be expressed as the KL divergence
\begin{equation}\label{eq:transfer-entropy2}
   TE_{s \to r} = \sum p(r_{t+1},r_t^{(m)},s_t^{(k)}) \log
   \frac{p(r_{t+1}|r_t^{(m)},s_t^{(k)})}{p(r_{t+1}|r_t^{(m)})} \,,
\end{equation}
where we define $s_t^{(k)} := (s_t,...,s_{t-k+1})$ and $r_t^{(m)} := (r_t,...,r_{t-m+1})$, which makes explicit that transfer entropy measures the log deviation from the generalized Markov property $p(r_{t+1}|r_t^{(m)}) = p(r_{t+1}|r_t^{(m)},s_t^{(k)})$. 
To address any non-stationarity of the sentiment processes, we perform an augmented Dickey–Fuller test to detect the presence of unit roots with p-values ($<0.01$) obtained through regression surface approximation\cite{mackinnon1994approximate, mackinnon2010critical}. We compute first differences of any unit roots processes we detect. 

Computationally, it is common  to set $m=k=1$, a convention we follow here.\cite{Schreiber_2000,EstimateInfo} Notably, there is no statistically significant (linear) autocorrelation in the price return series, supporting our choice of $m=1$. In this article, we use symbolic re-coding for calculation of transfer entropy, by partitioning the data into a finite number of bins based on the quantiles (5\%,25\%,50\%,75\%,95\%) of the empirical distribution. 
Transfer entropy with parameters $m=k\geq2$, was having numerical problems in estimating joint and conditional densities due to the small sample sizes of sentiment scores. Examining the exact ``speed'' of information absorption with $m=k\geq2$ is out of scope of our current work.

\subsection*{News sentiment model}
\begin{figure}[ht]
\centering
\includegraphics[width=1.0\linewidth]{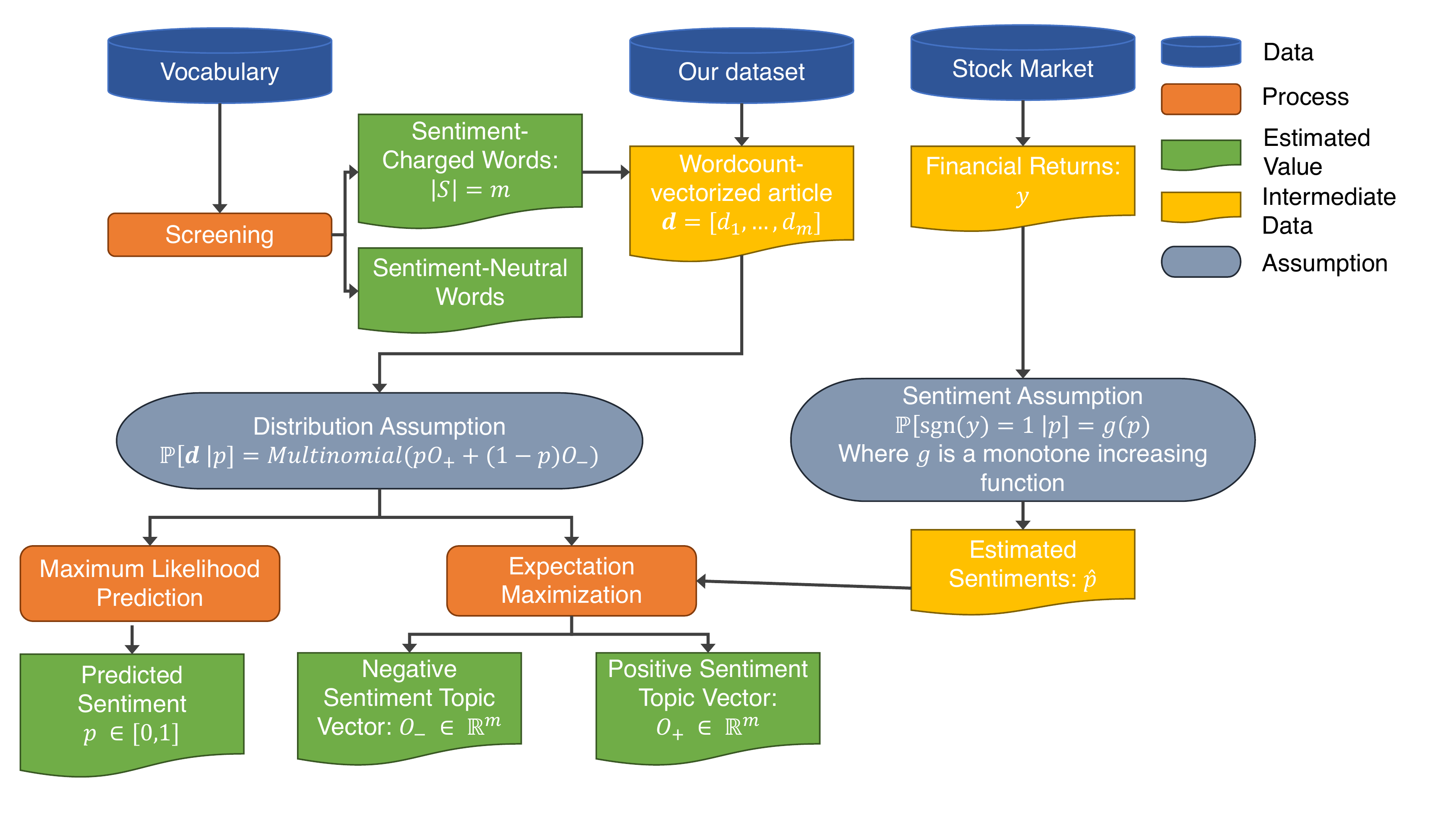}
\caption{Process chart of the sentiment model. The two assumptions underlying the sentiment model are depicted in light blue in the middle. The data used in fitting the model is shown at the top. We apply the predicted sentiment scores (bottom-left corner) to analyze transfer entropy and simulate several simple trading strategies.}
\label{fig:SESTM_model}
\end{figure}

In this study we use the Sentiment Extraction via Screening and Topic Modeling (SESTM) approach.\cite{kelly2019TextData} While closely related to the latent Dirichlet allocation (LDA) model\cite{10.5555/944919.944937} and vector-based representations such as word2Vec\cite{10.5555/2999792.2999959} and GloVe \cite{pennington-etal-2014-glove}, unlike these models SESTM is trained in a supervised fashion that facilitates interpretability and provides theoretical guarantees on the accuracy of estimates with minimal assumptions\cite{kelly2019TextData}.  Figure~\ref{fig:SESTM_model} depicts the main steps of the SESTM model we use in this article. We briefly describe our model below; a more detailed description can be found in the SI.

Before training the SESTM model, we apply standard pre-processing steps from NLP to turn articles into document-term vectors, including stopword-removal, tokenization and lemmatization. For each article $i$ in our dataset we assign a sentiment score $p_i \in [0,1]$ that reflects the financial sentiment the article bears towards the subject company, where $p_i = 0$ and $p=1$ denote the most negative and positive sentiment achievable, respectively. We view a sentiment of $p=0.5$ as neutral. We assume positive (negative) sentiment for a company most likely leads to positive (negative) return for the company's stock in the following sense
\begin{equation*}
    \mathbb{P}(\textrm{sgn}(r_i) = 1) = g(p_i) \,,
\end{equation*}
where $\mathbb{P}$ denotes the probability, $g(\cdot)$ is a monotonically increasing function and $r_i$ is the corresponding financial return of the company pertinent to news article $i$.

We assume that only a subset of the words in the corpus' dictionary is relevant and refer to these as \emph{sentiment-charged}. The remaining words are referred to as \emph{sentiment-neutral}. The model determines the sentiment-charged vocabulary of words, $S$, by including only those words that occur sufficiently frequently in our corpus and that are predominantly associated with either positive or negative returns. We remove all sentiment-neutral words from our original dictionary.

For each article we associate a document-term vector, $d_{i}$, of the occurrences of the sentiment-charged words and assume that it has a mixture multinomial distribution of the form
\[
    d_{i} \sim \textrm{Multinomial}\left(
        s_i, p_i O_+ + (1-p_i)O_-
    \right)
\]
where $s_i = \sum_{j \in S} d_{i,j}$ which scales the distribution while $p_i O_+ + (1-p_i)O_-$ is a mixture of two-topics that determines the probability distribution over the sentiment-charged words.  $O_+$ describes the probability of the words in a maximally positive article, $p_i = 1$. Similarly, $O_-$ describes that of a negative article, $p_i = 0$. We assume that $O_+, O_- \in \mathbb{R}^{|S|}_{+}$ are normalized such that $||O_+||_1 = ||O_-||_1 = 1$. For the articles with sentiment not on the boundary, $0 < p_i < 1$, word frequencies are convex combinations of those from the two topics. We train our model by estimating the vectors $O_+$ and $O_-$ via expectation maximization (EM), denoting their estimates by $\hat{O}_+$ and $\hat{O}_-$.

The sentiment $\hat{p}$ associated with a news article is determined by maximum likelihood estimation applied to the multinomial distribution of the new article's document-term vector, $d$. In other words, we determine $\hat{p}$ by solving
\[
    \hat{p} = \arg \max_{p\in [0,1]} \left\{
        \frac{1}{s}\sum_{j \in S}d_j\log\left(p\hat{O}_{j,+} + (1-p)\hat{O}_{j,-}\right)
        + \lambda \log (p(1-p))
    \right\} \,,
\]
where $s := \sum_{j \in S}d_j$ and $\lambda > 0$ is a regularization parameter. The choice of regularization is equivalent of imposing a beta prior on the sentiment, thereby pulling the estimated values toward the neutral score ($p=0.5$).

While  our assumptions are the same as the SESTM model \cite{kelly2019TextData}, we deviate from the original parameter estimation procedure due to the smaller size of our dataset. Specifically, the model parameters are estimated at the beginning of each month based on all the articles observed up until that point rather than estimating the parameters yearly based on the observations in the previous 15 years. 
Then the fitted model is used to predict sentiments during the whole month before being updated again. In addition, we keep the hyperparameters used for the sentiment-charged words selection and the prediction regularization parameter, $\lambda$, fixed during backtesting rather than considering them as a part of the periodic estimation. 

\bibliography{sample}

\section*{Acknowledgements}
F.F., M.J. and B.P thank the Professors Zhang Ce and Andreas Krause for their helpful comments and AWS credits during the Data Science Lab course. 
N.A.-F. acknowledges financial support from SoBigData++ through Grant Agreement No. 871042. M.J. acknowledges financial support from Public Scholarship and Development Fund of the Republic of Slovenia.

\section*{Author contributions statement}
F.F., M.J. and B.P. contributed equally to the work.
All authors contributed to the writing and editing of the manuscript. P.N.K. and N.A.F. designed and supervised this project. F.F., M.J. and B.P. processed the Common Crawl News data. All authors contributed to the modeling and analysis. In particular, F.E. did the exploratory data analysis, B.P. developed the research specific processing steps (the filtering of the financial news, company matching and removal of duplicates), B.P. and M.J. implemented the sentiment models, M.J. gathered and processed the financial data, M.J. and P.N.K. conducted the financial analysis (sentiment trading strategies and comparison to the Alexandria dataset), and N.A.F. performed the transfer entropy study.

\section*{Supplementary Information}

\begin{figure}[ht]
\centering
\includegraphics[width=\textwidth]{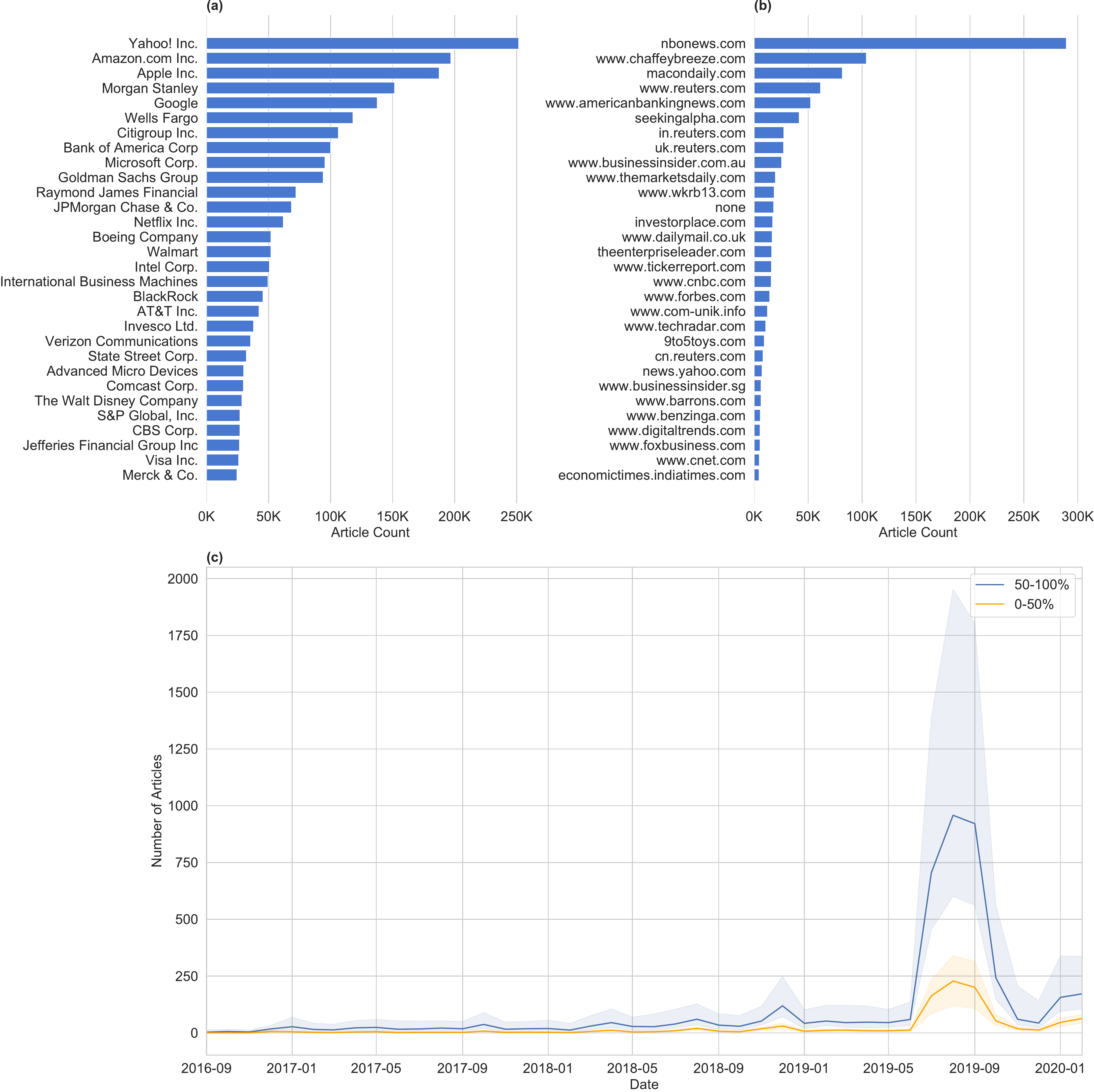}
\caption{Contrary to Figure \ref{fig:exploratory_analysis}, here we include NBO news. Summary of the Common Crawl News dataset. (a) Most frequently mentioned companies as measured by the number of distinct articles. (b) Most frequent news sources as measured by the number of distinct articles associated with each source. (c) Median number of articles published per company and month. The companies are divided into top and bottom halves by the total number of articles published about them. The shaded regions represent the 25\% and 75\% percentiles of each half.}
\label{fig:exploratory_analysis_SI}
\end{figure}

\subsection*{Our Common Crawl financial news dataset}
We provide open-source access to the financial news dataset from the Common Crawl News that we extracted and processed as part of our study. The dataset is available \href{https://drive.google.com/drive/folders/12q0uZ7D_0mb1GHqf1pTCAwDagbEnSALz?usp=sharing}{online}.\footnote{\url{https://drive.google.com/drive/folders/12q0uZ7D_0mb1GHqf1pTCAwDagbEnSALz?usp=sharing}} It consists of more than a million articles of financial news. The data processing pipeline we deployed is depicted in the Figure~\ref{fig:process_pipeline} and described in the section ``Common Crawl data processing'' in the main article. The fields available in the dataset are:
\begin{itemize}
    \item \textbf{publish\_date\_utc}: publishing date and time of the article, UTC time zone, 
    \item \textbf{crawl\_time\_utc}: crawling date and time of the article, UTC time zone,
    \item \textbf{url}: URL address from where the news was obtained,
    \item \textbf{title}: title of the news article, 
    \item \textbf{text}: body of the news article, 
    \item \textbf{mentions\_counter}: list containing all S\&P 500 companies mentioned in the article and the number of mentions of each company,
    \item \textbf{company}: name of the  company most pertinently aligned with the article based on our company matching model (for its description, see section ``Common Crawl data processing'' in the main article), and 
    \item \textbf{ticker}: ticker symbol of the company that was aligned with the article.

Note: The data is made available ``as is'' for general research purposes. We take no responsibility for the use of the data and we are not able to provide technical support for the data. 

\end{itemize}

\subsection*{Common Crawl data processing}
We use a four stage process to clean and transform the data from Common Crawl: (i) pre-screening, (ii) filtering, (iii) alignment and (iv) pre-Processing. Figure~\ref{fig:process_pipeline} in the main article depicts this process.

\textbf{Pre-screening.}
The first stage, referred to as \emph{pre-screening}, parses all the articles crawled in the dataset, extracts the relevant information (title, main text, URL, language, publication time) from the raw HTML strings\footnote{We used the Goose Python package, \href{https://github.com/goose3/goose3}{https://github.com/goose3/goose3}, to parse the HTML strings.} and saves only the articles that are written in English, have publication timestamp available and mentions a selected company at least once. In order to identify the constituent companies in the S\&P 500, we manually curated a set of keywords for each company. For example, we use the keywords ``IBM'' and ``International Business Machines'' for the company International Business Machines Corporation. We map each article to a list of companies with the number of keyword appearances.

Parsing is the most computationally expensive task in this stage especially extracting the main text. However, the language and publication time of an HTML file can be found easily with the right HTML tags. Furthermore, searching for keywords is also less expensive than parsing, therefore, we applied the language and publication time filter followed by a keyword filter, i.e., keeping articles with at least one keyword appearance, on the raw HTML string. With this we could reduce the number of articles and speed up the process. Once the parsing has been done, we applied the aforementioned steps to extract keywords that are present only in the main text and double check our previous language and publication time extraction. We deployed a Spark cluster in the same AWS region where the CommonCrawl data is located \footnote{The CommonCrawl data is located on the us-east-1 AWS region.} to minimize data transfer. This stage reduces the data from roughly 7.5TB to around 8GB, and therefore the proceeding stages can be ran on a single machine. 

\textbf{Filtering.}
The second stage, referred to as \emph{filtering}, parses the list of companies for each article from the pre-screening stage and removes the companies whose keywords appear only once in the text. We decided to omit companies that occur only once because they are unlikely to be the focus of the article. This stage also removes articles that do not have a UTC publication timestamp. The last filter in this stage is a random-forest based classifier that separates financial and non-financial news,  only keeping the first category. The details of this model are discussed in the following section.

\textbf{Alignment.}
After the previous steps, we have a dataset of financial articles written in English with all the additional information listed in the section ``Our Common Crawl financial news dataset.'' Naturally, companies can be references in an article for several reasons; an occurrence does not immediately imply that referenced company is the main subject of the underlying article. The purpose of the third stage of the pipeline, referred to as \emph{alignment}, is to determine the main subject company  of the article. We made the assumption that the main subject company of an article should be referenced early in the text. For instance, the company may occur in the title or in the first few paragraphs. Notably, the Common Crawl News dataset contains new articles from several sources which have duplicated content. We found that due to news aggregators some articles are reported by multiple domains. Specifically, we found that about 166K distinct titles appeared more than once, totalling about 450K duplicate entries. 
We removed duplicate entries by matching the titles and by approximate matching of the article body.

\textbf{Pre-processing.}
The last stage, referred to as \emph{pre-processing}, prepares the data for the sentiment model. First, it applies a standard NLP pipeline using the NLTK package\cite{BirdKleinLoper09} consisting of POS-tagging; proper noun removal; conversion of all text to lower case; deletion of special characters, English stopwords and non-English words; and lemmatization, Second, all the articles are countvectorized. Third, we calculated the open-to-open returns from day $t-2$ to $t+1$ around the publishing time $t$. The returns were added to our dataset as labels to be used for the estimation of our supervised sentiment model.

\subsubsection*{Financial filter model}
    As part of the filtering stage of the data processing pipeline, we trained a random forest classifier to distinguish between financial and non-financial articles. The procedure starts with countvectorization that reduces the tokens to a vocabulary learned on the training data. We obtained our training dataset from webhose.io \footnote{The dataset is available from \href{https://webhose.io}{https://webhose.io}},  consisting of English news articles categorized into seven categories (Entertainment, Travel, World, Technology, Political, Sports, Finance). We take into consideration that some processing steps have already been applied on the CommonCrawl dataset before the financial filter, namely, language, timestamp and company mentioning checks. We aim to train the financial filter model on a dataset as similar to the current stage of the main dataset as possible, therefore, we make the same checks. In particular, the language of the articles in the Webhose.io dataset is English filtered by the source, the timestamp is irrelevant in this case but we apply the company mention check and keep only the articles in which a company from our keyword dictionary appears. The resulting training data consists of 67599 samples out of which 14283 were in the Finance category. We assigned a positive label to the Finance category and a negative label for all the other categories. With cross-validated parameters, the model achieved 86\% accuracy, 70\% recall and 8\% false positive rate with the standard decision threshold parameter\footnote{The decision threshold parameter is the probability value above which an article is considered to be part of the positive category.} of 0.5 on a 80-20 train-test split of the data. The company alignment stage of the data processing pipeline cleans further our dataset, therefore, false positive predictions are more tolerable than false negatives. Accordingly, we considered lowering the decision threshold parameter to achieve higher recall. We evaluated the trained model on three other datasets that were not used for training: the Financial News Dataset\footnote{The data is available from \href{https://github.com/duynht/financial-news-dataset}{https://github.com/duynht/financial-news-dataset}}, the Reuters Datasets\footnote{The data is available from \href{https://figshare.com/projects/Financial-News-Full-Details-JSON-Dataset/23893}{https://figshare.com/projects/Financial-News-Full-Details-JSON-Dataset/23893}} and the BBC Dataset \footnote{The data is available from \href{https://storage.googleapis.com/dataset-uploader/bbc/bbc-text.csv}{https://storage.googleapis.com/dataset-uploader/bbc/bbc-text.csv}}. Table~\ref{tab:recall_table} shows the recall of the model achieved for different test datasets and decision thresholds. 
    \begin{table}[ht]
    \centering
    \begin{tabular}{|l|c|c|c|c|}
    \hline
     Decision Threshold & Webhose.io & Financial News & Reuters & BBC News \\
    \hline
    0.3  & 97\% & 99\% & 98\% & 95\% \\
    \hline
    0.35 & 89\% & 99\% & 97\% & 90\% \\
    \hline
    0.4  & 84\% & 98\% & 90\% & 89\% \\
    \hline
    0.5  & 70\% & 94\% & 78\% & 62\% \\
    \hline
    \end{tabular}
    \caption{\label{tab:recall_table} Recall values obtained from the evaluation of the financial filter model. The first column specifies the decision threshold, i.e., the probability value above which an article is considered to be part of the positive category.}
\end{table}

\subsection*{News sentiment model}
In this section, we describe the sentiment model in more details. Our model assumptions and estimation procedure for a \emph{fixed dataset} is the same as the SESTM model in Ke et al.\cite{kelly2019TextData}. However, we deviated from Ke et al. in how we  recalibrated the model over time. 

First, we establish the notation used in this section. Let $n$ denote the number of news articles, $m$ the size of the dictionary, $d_{ij} \in \mathbb{Z}_{\geq 0}$ the number of times word $j$ appears in article $i$ and $d_i \in \mathbb{Z}_{\geq 0}^m$ the vector of all the word counts corresponding to the $i$-th article. We form a $n \times m$ matrix, $D$, by concatenating the vectors $\{d_i\}_{i=1}^n$ row-wise. In some cases, we will take only a subset of the dictionary, denoted by $S$, then $d_{i, [S]}$ will denote the subset of the vector $d_{i}$ consisting of entries only from the subset $S$. A financial label corresponding to article $i$ is denoted by $r_i$.

The main assumption of the model is that each article $i$ possesses a sentiment score, $p_i \in [0,1]$, where $p_i = 0$ is the most negative and $p_i = 1$ the most positive. We view $p_i = 0.5$ as the neutral value. The sentiment score serves as a sufficient statistic for the document's countvector and financial return, i.e., $d_i$ and $r_i$ are independent given $p_i$. Furthermore, we assume that $\mathbb{P}(sgn(r_i) = 1) = g(p_i)$ where $\mathbb{P}$ denotes the probability and $g(\cdot)$ is a monotonically increasing function and the dictionary has a disjoint partition, $\{1, \dots, m \} = S \cup N$, where $S$ is the set of \textit{sentiment-charged} words and $N$ is the set of \textit{sentiment-neutral} words. We assume that the sentiment-charged count vector, $d_{i,[S]}$, and the sentiment-neutral count vector, $d_{i,[N]}$, are independent of each other and the latter is essentially a nuisance, and due to its independence from the vector of interest, $d_{i,[S]}$, the modeling of $d_{i,[N]}$ is omitted. \footnote{In the main text $d_i$ denotes $d_{i, [S]}$ to simplify our notation.} The sentiment model builds on the assumption of the sentiment-charged word vector's Multinomial distribution, i.e,.
\[
    d_{i} \sim \textrm{Multinomial}\left(
        s_i, p_i O_+ + (1-p_i)O_-
    \right)
\]
where $s_i = \sum_{j \in S} d_{i,j}$ which scales the distribution while $p_i O_+ + (1-p_i)O_-$ is a mixture of two-topics that determines the probability distribution over the sentiment-charged words.  $O_+$ describes the probability of the words in a maximally positive article, $p_i = 1$, and similarly $O_-$ that of a negative article, $p_i = 0$. We assume that $O_+, O_- \in \mathbb{R}^{|S|}_{\geq 0}$ are normalized such that $||O_+||_1 = ||O_-||_1 = 1$. For the articles with sentiment not on the boundary, $0 < p_i < 1$, word frequencies are convex combinations of those from the two topics.

\subsubsection*{Determining sentiment-charged words}
The model determines the sentiment-charged vocabulary of words, $S$, by including those words that occur sufficiently frequently in our corpus and that are predominantly associated with either positive or negative returns. Define $k_j = \#\{\textrm{Articles including word } j\}$ and
$
    f_j = \frac{
        \#\{\textrm{Articles including word $j$ AND having $\textrm{sgn}(y_i) = 1$}\}
    }{
        k_j
    }
$, then we estimate $S$ with
\[
    \hat{S} = (\{j: f_j > 1/2 + \alpha_+\}\cup\{j: f_j < 1/2 - \alpha_-\})\cap \{j: k_j > \kappa\}
\]
where $\alpha_+, \alpha_-$ and $\kappa$ are parameters of the model. This procedure is referred to as marginal screening in the statistical literature \cite{fan2008sure} and its "sure screening" property has been established \cite{kelly2019TextData}, i.e., $\mathbb{P}(S = \hat{S}) \to 1$ as $n\to \infty$ and $m \to \infty$.

\subsubsection*{Estimating the topic vectors}
The model estimates the vectors $O_+$ and $O_-$ via expectation maximization. Since both of these vectors and the $p_i$ values are parameters of the Multinomial distribution over $d_{i,[S]}$, we first assume that the values $p_i, i = 1,...,n$ are known. Let $\tilde{d}_{i,[S]} = d_{i,[S]} / s_i$ denote the vector of word frequencies where $s_i = \sum_{j \in S} d_{i,j}$, then the Multinomial distribution implies that
\[
    \mathbb{E}[\tilde{d}_{i,[S]}] = p_i O_+ + (1-p)O_-
\]
In order to estimate $O = [O_+, O_-]$ via this formula, we need the values of $p_1,...,p_n$ and $S$ which are not directly observable. So, we use $\hat{S}$ to estimate $S$ and approximate $p_i$ as
$
    \hat{p}_i = \frac{\textrm{Rank of $y_i$ in } \{y_i\}_{i=1}^n}{n}
$.
The estimation is then done via expectation maximization that in this case has the following closed form solution
\[
    \hat{O} = \Tilde{D}^\intercal \hat{W}^\intercal (\hat{W}\hat{W}^\intercal)^{-1}
\]
where
\[
    \hat{W} = \begin{bmatrix}
        \hat{p}_1 & \dots & \hat{p}_n \\
        1-\hat{p}_1 & \dots & 1-\hat{p}_n
    \end{bmatrix}
    \textrm{ and }
    \tilde{D} = \begin{bmatrix}
        \tilde{d}_1 & \tilde{d_1} & \dots & \tilde{d_n}
    \end{bmatrix}^\intercal
\]

\subsubsection*{Calculating sentiment for new articles}
For new articles, we estimate the underlying sentiment value $p$ with the maximum likelihood estimation procedure applied on the multinomial distribution of $d_{i,[S]}$ i.e.
\[
    \hat{p} = \arg \max_{p\in [0,1]} \left\{
        \frac{1}{\hat{s}}\sum_{j \in \hat{S}}d_j\log\left(p\hat{O}_{j,+} + (1-p)\hat{O}_{j,-}\right)
        + \lambda \log (p(1-p))
    \right\}
\]
where $\hat{s} = \sum_{j \in \hat{S}}d_j$. An additional penalization term is added with parameter $\lambda > 0$. This is equivalent of imposing a Beta prior on the sentiment and it pulls the estimated values toward the neutral $0.5$ score.

\subsubsection*{Rolling-time Parameter Estimation}
The model's original fitting procedure estimates the parameters of the SESTM model on a 15 year time-window and use the estimated values for prediction in the subsequent one-year period. This procedure is not suitable to our smaller dataset, therefore we refitted the model's parameters at the beginning of each month using all the articles prior to that date. Each estimated model was then used to predict new articles until the next recalibration a month later. The data between August 26, 2016 and January 1, 2018 have been used as a warmup period where we did not conduct any trading. Additionally, the procedure in \cite{kelly2019TextData} changes the hyperparameters of the model each year based on a 10-5 year training-validation split of the 15 years window. Due to our smaller dataset, we could not implement this parameter selection heuristic and used a fixed set of hyperparameters.

\subsection*{Choice of sentiment model}
To explore the effect of the choice of sentiment model on the information analysis, we also looked at finBERT \cite{araci2019finbert}, which was shown to achieve state-of-the-art performance in several financial sentiment tasks. It is a language model based on BERT\cite{devlin2018bert} and fine-tuned on financial data. While SESTM \cite{kelly2019TextData} model falls within more traditional statistical family of models (based on count statistics and bag-of-words features), BERT builds upon the latest deep learning advancements and is well known within NLP community. \par
Due to the much higher computational load for sentiment prediction using finBERT, we considered here shorter investment period from 3rd of January 2018 to 31st of December 2018. There are around 220 thousand news to trade on for this period in our parsed Common Crawl dataset, so sample size is still representative enough. We used finBERT off-the-shelf, i.e. no additional training was done on Common Crawl data. For each sentence finBERT returns the probability of having positive, neutral and negative. We aggregated these to obtain scores for each news as:
$$
\textrm{news sentiment}:= \frac{\# \textrm{ positive sentences }- \# \textrm{ negative sentences}}{\# \textrm{ sentences}}
$$
\par
\begin{figure}[ht]
\centering
\includegraphics[width=1.0\linewidth]{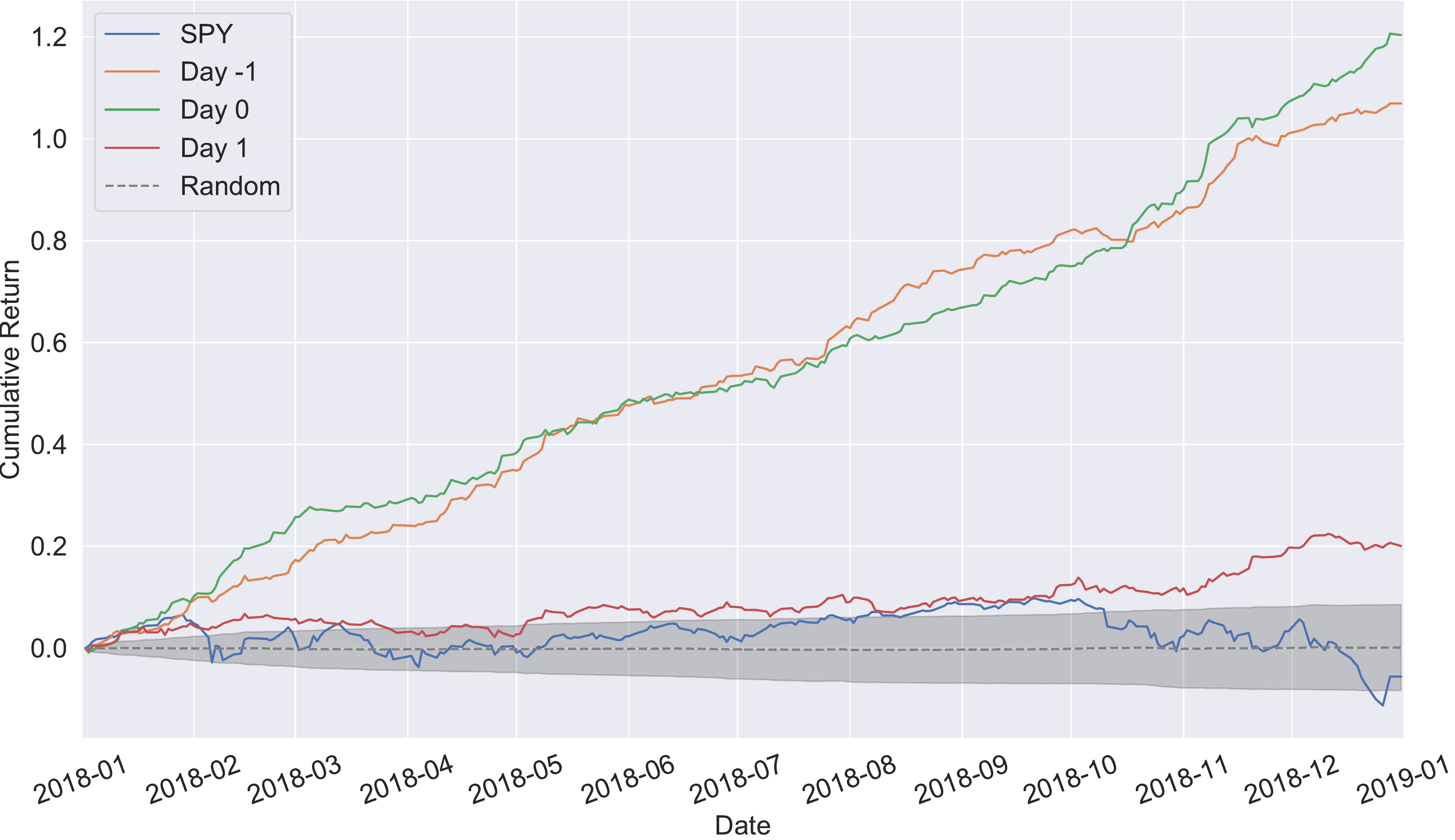}
\caption{Cumulative returns of trading strategies and benchmarks for sentiment extracted with finBERT. ``Day 1'' represents the cumulative returns of the Day 1 sentiment strategy based on the Common Crawl dataset from January 2018 through December 2018. SPY is the SPDR S\&P 500 trust. ``Random'' denotes the average of the random strategies along with one standard deviation confidence bands obtained from 500 simulations. ``Day 0'' and ``Day -1'' are the ``look-ahead'' sentiment strategies relying on future information.}
\label{fig:trading_results_BERT}
\end{figure}

Based on results in Figure \ref{fig:trading_results_BERT} and Table \ref{tab:trading_results_BERT} we conclude that stronger signal could be found in Common Crawl dataset by deploying (and carefully tuning) more sophisticated sentiment models like finBERT. However due to computational constraints and since we deemed SESTM results satisfactory for the purposes of our analysis we did not look into it further. Nevertheless we note, that this could be a fruitful area of future work, which could lead to an even better understanding of the dynamics between public information on the World Wide Web and financial markets.

\begin{table}[ht]
\setlength{\tabcolsep}{7pt}
\centering
  \begin{tabular}{llllll}
    \toprule
    \multirow{5}{*}{} &
      \multicolumn{3}{c}{\textbf{Common Crawl (finBERT)}}  &
      \multicolumn{1}{c}{} &
      \multicolumn{1}{c}{}\\
      & {Day -1} & {Day 0 } & {Day 1} & {\textbf{SPY}} & {\textbf{Random}}  \\
      \midrule
    Ann. avg. return & 107.36\%  & 120.83\% & 20.13\%  & -5.60\% & $ 0.11\pm 8.46\%$ \\[0.1cm]
    Ann. volatility &  11.19\%  &  11.84\% &  9.92\%  &  17.30\% & $ 8.17\pm 0.45\%$  \\[0.1cm]
    Ann. Sharpe ratio (p-value) & 9.60 (< 0.001) & 10.21 (< 0.001) & 2.03 (< 0.001)  & -0.32 & $ 0.01 \pm 1.04 $ \\[0.1cm]
    MDD & 2.63\%  & 1.8\% & 4.6\%  & 21.04\% & $ 9.54\pm4.46\% $  \\[0.1cm]
    
    Ann. $\alpha$ (p-value) & 108.36\% (< 0.001)  & 120.96\% (< 0.001) & 20.16\% (0.044)  & 0  & $ 0.46 \pm 8.50\% $\\[0.1cm]
    $R^2$ & 0.015  & 0.027 &  0.001 & 1 & $ 0.006 \pm 0.008 $ \\
    \bottomrule
  \end{tabular}
  \caption{\label{tab:trading_results_BERT} Performance statistics of the Day 1 sentiment trading strategy and benchmarks from 3rd of January 2018 through 31st of December 2018. The sentiment trading strategy is based on news articles from the Common Crawl News dataset and sentiment is extracted using finBERT. SPY is the SPDR S\&P 500 trust. ``Random'' denotes the baseline strategy where each day we randomly select companies to invest in. ``Day 0'' and ``Day -1'' are ``look-ahead'' sentiment strategies, reported for comparison purposes. Statistics are computed using daily returns (n=251). MDD is the Maximum Daily Drawdown defined as the maximum observed decline from a historical peak of the price until a new peak is attained. The p-values for the Sharpe ratios were bootstrapped from 500 random backtests. We obtain $\alpha$ (the intercept) and $R^2$ by regressing the daily returns of the portfolios on the daily returns of the SPY. Similarly, the performance metrics of the random portfolios were bootstrapped from 500 random backtests.}
\end{table}

\subsection*{Comparing the information transfer of  private and public news to financial markets}
  
 Here we report some additional results on comparison between Alexandria and Common Crawl datasets. \par
 Through our experiments we have observed that using only simple trading configuration (rebalancing daily at market open using all news from last trading day's market open on) does not explain the whole picture. Hence, to better understand the difference between the two datasets, we include two other portfolio rebalancing approaches. For the first approach, we rebalance the portfolio weekly by using all news published during the prior week to determine sentiment scores. While on the one hand, using a longer horizon to determine the news sentiment for each company, we have more articles to rely on. On the other, some of these articles might be “stale” and therefore no longer relevant. For the second approach, we rebalance the portfolio daily by using only news published outside of market open hours, i.e. after 4PM of the last trading day. This way we are excluding all news that market participants could act on already during market open hours of the last trading day and are hence making investment decision on the current trading day using only news that were not traded on already. We note that we neglect the possibility to trade outside of market open hours here. While this surely is a simplification, we would argue it is not a strong one, as trading volumes are much lower outside of market open hours. \par
 In Table \ref{tab:trading_results_diff_config_new} we report trading results. Based on simple daily regime (denoted by \emph{Daily-9.30AM}) we detect similar amount of financial information present in both datasets as measured by Sharpe ratio: 1.64 for CommonCrawl and 1.51 for Alexandria. However, concluding that there are no big differences between the two datasets based on this finding would be premature, as illustrated by trading metrics for the other two regimes. For weekly rebalancing, Common Crawl results slightly improve (Sharpe 1.69), while Alexandria results deteriorate as indicated by negative Sharpe ratio (-0.755). On the other hand, excluding intraday news when constructing sentiment investment portfolio (\textit{Day $-$ 4.00PM} rebalancing) hurts Common Crawl performance (Sharpe 1.08), while Alexandria results improve, achieving best overall trading performance (Sharpe 2.46).
 
 These results suggest that Common Crawl captures more long-term trend, whereas Alexandria seems to be calibrated to be more 'fast-moving' and geared towards intraday trading. Differences when varying the trading rebalancing rules can be partly explained by Common Crawl being an alternative dataset and not as widely used by market participants as Alexandria, hence inclusion of stale news improves rather than hurts the trading performance. The difference in information can also be attributed to the different sentiment models used to extract signals. While for Common Crawl sentiment SESTM \cite{kelly2019TextData} model is used, Alexandria's model is proprietary and we did not get access to their methodology.
 
\begin{figure}[ht]
\centering
\includegraphics[width=1.0\linewidth]{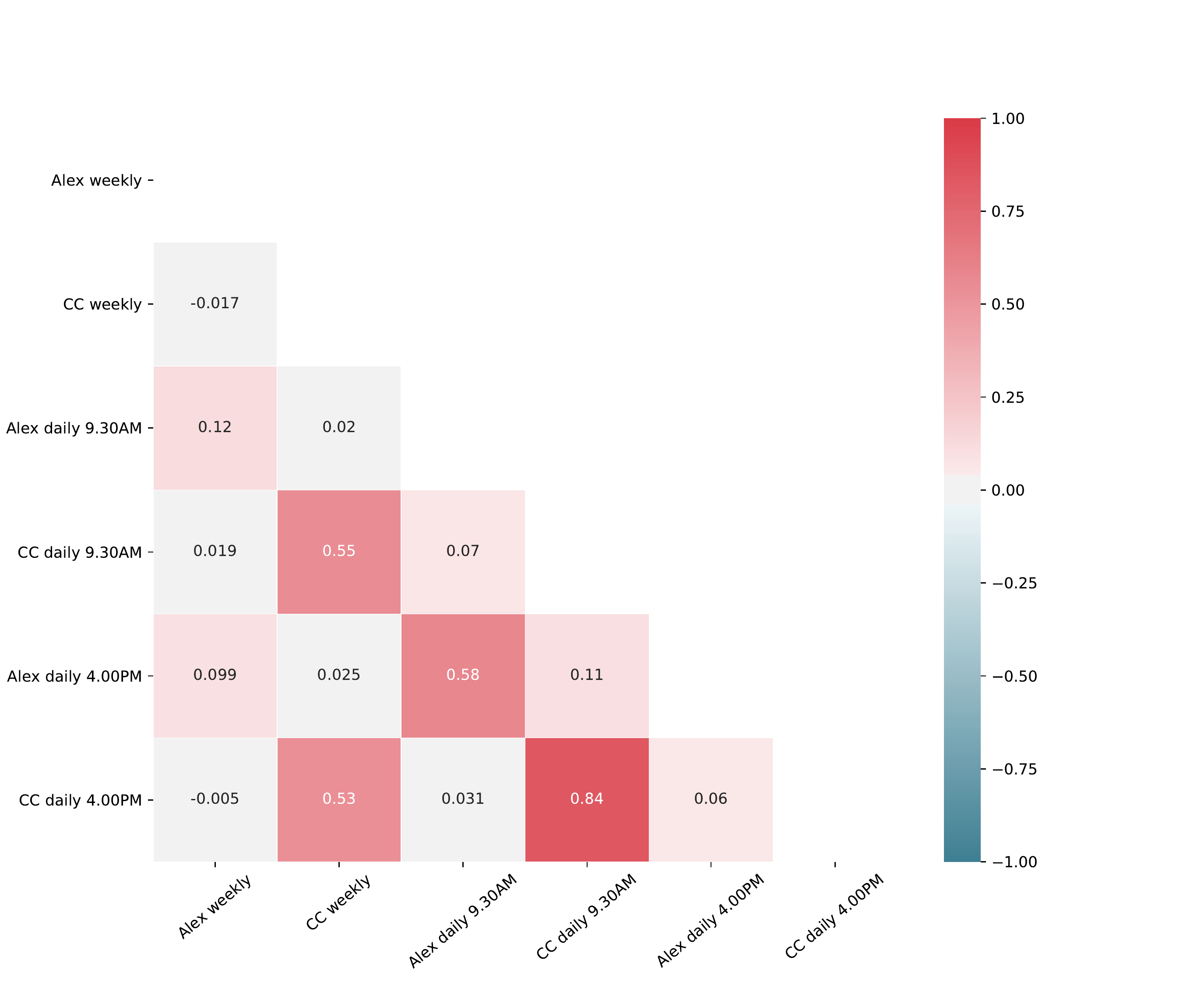}
\caption{Correlation matrix of all simulated trading strategies using the  Alexandria and Common Crawl News datasets. Each trading strategy has $n=542$ daily returns.} 
\label{fig:corr_mat} 
\end{figure}

 The difference between information captured in the two datasets is also additionally confirmed with low correlation of 0.07 (statistically not different than 0 with p-value for one sided t-test of 0.105) between \emph{Daily-9.30AM} rebalancing of Alexandria and Common Crawl. We report correlations for all other simulations in the Figure \ref{fig:corr_mat}.
 
 \begin{table}[ht]
\setlength{\tabcolsep}{8pt}
\centering
  \begin{tabular}{llll|lll}
    \toprule
    \multirow{5}{*}{} &
      \multicolumn{3}{c}{\textbf{Common Crawl}}  &
      \multicolumn{3}{c}{\textbf{Alexandria}} \\
      & {Weekly} & {Daily-9.30AM} & {Daily-4.00PM} & {Weekly} & {Daily-9.30AM} & {Daily-4.00PM}  \\
      \midrule
    Ann. avg. return & 23.30\%  & 21.02\% & 13.94\%  & -6.15\% & 14.87\% & 23.24\% \\[0.1cm]
    Ann. volatility &  13.77\%  &  12.97\% &  12.85\%  &  8.16\% & 9.84\% & 9.47\% \\[0.1cm]
    Ann. Sharpe ratio & 1.69  & 1.64  & 1.08   & -0.755 & 1.51  & 2.46 \\[0.1cm]
    MDD & 12.41\%  & 10.31\% & 11.10\%  & 20.16\% & 7.27\%  & 7.87\% \\[0.1cm]
    
    Ann. $\alpha$  & 23.26\%   & 20.69\%  & 13.38\%   & -5.83\%  & 15.22\% & 23.58\% \\[0.1cm]
    $R^2$ &  0.0002 & 0.0038 &  0.0093 & 0.0078 & 0.005 & 0.0093\\
    \bottomrule
  \end{tabular}
  \caption{\label{tab:trading_results_diff_config_new} Trading performance statistics for different rebalancing rules based on sentiment scores from the Alexandria and Common Crawl News datasets from January 2018 through February 2020. Denoting current trading day by $day_t$, time intervals from which we consider news for portfolio construction are 9:30 a.m. on day ${t-5}$ through 9:00 a.m. on day $t$ for \textit{Weekly} rebalancing, 9:30 a.m. on day ${t-1}$ through 9:00 a.m. on day $t$ for \textit{Daily-9.30AM} rebalancing and 4:00 p.m. on day ${t-1}$ through 9:00 a.m. on day $t$ for \textit{Daily-4.00PM} rebalancing. In all simulations, we enter positions at 9.30 a.m. on day $t$. For \textit{Weekly} and \textit{Daily} rebalancing we hold positions until 9.30 a.m. on day $t+5$ and 9.30 a.m. on day $t+1$, respectively. Statistics are computed using daily returns (n=542). MDD is the Maximum Daily Drawdown defined as the maximum observed decline from a historical peak of the price until a new peak is attained. We obtain $\alpha$ (the intercept) and $R^2$ by regressing the daily returns of the portfolios on the daily returns of the SPY.}
\end{table}

\subsection*{Financial market data}
For daily returns of S\&P500 companies we used \emph{Yahoo Finance} \footnote{\href{https://finance.yahoo.com}{https://finance.yahoo.com}} price data. Since our simple trading strategy consists of entering or closing positions at the opening of market, open prices were used for backtesting purposes. Financial labels in SESTM\cite{kelly2019TextData} model were calculated based on adjusted close daily prices from Yahoo Finance. For intraday returns we worked with \emph{Wharton Research Data Services} (WRDS) \footnote{\href{https://wrds-www.wharton.upenn.edu}{https://wrds-www.wharton.upenn.edu}}, specifically with National Best Bid and Offer (NBBO) dataset. We aggregated reported milisecond quotes on a minute level and used last bid and last ask to compute mid-quote prices for each minute.

\subsection*{Extended related work}
The related work on sentiment analysis is vast, due to the large history in finance, multidisciplinary approaches, and technological advancements in NLP and ML research.
Making exhaustive review \cite{SentRev1,mantyla2018evolution,SentFinEvo,ain2017sentiment} on the use of sentiment analysis in finance is out of the scope of this work. However, we provide a short overview to the research, by dividing it, based on the source of alternative data.

\textbf{Social media data.}
The effects of collective mood changes have been researched exhaustively over time. Most of the papers consider Twitter sentiments as an indicator.  
Mao et al. (2011) \cite{mao2011predicting} uses OpinionFinder and Google-Profile of Mood States (GPOMS) to track sentiment and other dimensions (calmness, alertness, sureness, vitalness, kindness, happiness) of daily Twitter feeds. Shows that these values are indeed be used to tracking public mood state. Predictive power analysis of DJIA closing prices with Granger Causality and Self-Organizing Fuzzy Neural Network. Only some of the variables e.g. calmness are useful. Focusing on increase in Mean-Average-Percentage-Error of a baseline model if sentiment added and predicting only direction but not magnitude.
Rao et al. (2012) \cite{rao2012analyzing} analyses tweet sentiments' capability to forecast financial returns, volatility, volume and close prices of large cap US stocks. They found that sentiments, extracted via Standford NLP group's algorithm, have large correlation with the investigated dependent variables. Their analysis shows that the optimal time window (look-back time from which features are generates) is one month.
Zheludev et. al (2014) \cite{zheludev2014can} use sentiment analysis methods and information theory measures to show that social media message sentiment can contain information on the future prices of the S\&P500 index. 
Ranco et al. (2015) \cite{ranco2015effects} found that twitter volume and sentiment information have significant dependencies with market returns.
Broadstock et al. (2019) \cite{broadstock2019social} demonstrate that sentiment from social media contains pricing power against the stock market.

\textbf{Web data.}
Wang et al. (2013) \cite{wang2013financial} analyze the relation between financial sentiment (lexicon-based) and risk (volatility of returns).
Schumaker et al. (2013) \cite{schumaker2009textual} consider different textual representations (bag of words, noun phrases, named entities) to fit a SVM model and predict stock prices 20 minutes after an article release.
Bordino et al. (2014) \cite{bordino2014stock} study relationship between Yahoo web traffic and trading volumes. Use time-lagged cross-correlation to show dependence between web traffic and trading volumes. Find correlation over two to three days, and find that correlation increases when moving from hourly granularity to daily granularity.
Piskorec et al. (2014) \cite{piskorec2014cohesiveness} propose a measure of collective behaviour based on financial news on the Web, the News Cohesiveness Index (NCI) and quantify the relation to a market volatility.
Zhang et al. (2014) \cite{zhang2014internet} analyze information arrival of financial news with a proxy of World Wide Web News to characterize market volatility. 
Curme et al. (2014) \cite{curme2015coupled} analyze the network structure of lagged correlations among financial news sentiment and price returns. 
Alanyali et al. (2015) \cite{alanyali2013quantifying} analyze the statistical relationships between the company mentions in financial news and daily transaction volumes. 
Rieis et al. (2015) \cite{dos2015breaking} Analyze how to come up with the headline (of a news, article etc.) that will attract a lot of clicks online, they also investigate an effect of clickbaits. They show that extremely negative and positive headlines tend to attract more popularity.
Heiberger (2015) \cite{heiberger2015collective} looks at relationship between Google search volumes and stock prices. Trading strategy based on google search volumes beats the market during financial turmoil. Break down results by company and by sector.
Ranco et al. (2016) \cite{lilloYahoo2016} look at relationship between news sentiment and stock returns. In addition, weigh news article based on number of clicks. Use correlation and Granger causality to show relationship between click-weighted sentiment and stock returns.
ang et al. (2017) \cite{yang2017genetic} propose a trading strategy based on the sentiment feedback strength between the news and tweets using generic programming optimization method. They investigate relationship between tweets sentiment, news sentiment and market returns. They work with Northern Light SinglePoint business news portal data and their sentiment model is lexicon-based.
Amin et al. (2019) \cite{amin2019sentiment} analyze abstracts of financial news with sentiment methods to predict the market trends.
Choudhury et al. (2008) \cite{de2008can} and Ruiz et al. (2012) \cite{ruiz2012correlating} analyze statistical relationships between micro-blogging activities and stock market variables.

\textbf{Search queries.}
Bordino et al. (2012) \cite{bordino2012web} analyze the statistical relationship between trading volumes in NADAQ-100 and daily search queries on Yahoo.
Preis et al. (2013) \cite{preis2013quantifying} analyze changes in Google query volumes related to finance to extract early warning indicators of stock market moves. 
Zhang et al. (2013) \cite{zhang2013open} propose to use the proxy for investor attention by employing the search frequency of  stock information in Baidu World Wide Web search queries. Quantify the effect of investor attention proxy with abnormal return with Granger causality methodology.

\textbf{Common Crawl Web archives data analysis.}
Du et al. (2017) \cite{du2017representativeness} studies representativeness of Common Crawl data using topic models. They find that proportions in Common Crawl data not significantly different from topic proportions in full web data.
Wang et al. (2019) \cite{wang2019construction} build an ensemble of deep learning models (CNN, Att-BLSTM and BERT) for prediction of sentiment of financial news. They focus on the sentiment model, not on the datasets. They do not work with Common Crawl dataset, only one of the models in their ensemble was pre-trained on part of Common Crawl.
Mehmood et al. (2017) \cite{mehmood2017understanding}  study language distribution across the web using Common Crawl data.
Schelter et al. (2018) \cite{schelter2018ubiquity} perform large-scale analysis of third-party trackers on the World Wide Web using Common Crawl billion-page web data.

{\bfseries Financial economics.}
Additionally, it is worth pointing to the work in the area of financial economics and sentiment analysis. 
Chan (2003) \cite{chan2003stock} analyzed the database of headlines about individual companies and monthly returns and compared it to the stocks with similar returns, but no identifiable public news.
Maheu et al. (2004) \cite{maheu2004news} models the conditional variance of price returns as a combination of jumps and smoothly changing components. With this model they have empirically improved volatility forecasts.  
Vega (2006) \cite{vega2006stock} analyzed private information trading variables and public news databases in order to quantify market impact. 
In this work it has been identified the important factor is whether information is associated with the arrival rate of informed or uninformed traders
Tetlock (2007) \cite{tetlock2007giving} analyzed the nature of media's interaction from Wall Street Journal column and stock market. It was found that high media pessimism predicts downward pressure on market prices and unusually high or low pessimism predicts high
market trading volume.
Gro{\ss}-Klu{\ss}mann et al. (2011) \cite{gross2011machines} studied high-frequency market reactions to an intraday stock-specific news flow and found sentiment indicators have predictability for future price trends but for the profitability bid-ask spreads interactions are needed. 
Birz et al. (2011) \cite{birz2011effect} analyzed macroeconomic news to better estimate its effect on stock returns and found that news about GDP and unemployment does affect stock returns.
Engelberg et al. (2012) \cite{engelberg2012shorts} analyzed database of short sales combined with a database of news releases and found that a substantial portion of short sellers' trading advantage comes from their ability to analyze publicly available information.
Hisano et al. (2013) \cite{hisano_sornette2013} analyze the news from Thompson Reuters and relationship with trading activity of 206 major stocks in S\&P US stock index. They provide statistical evidence for effect of the news flow to explain “abnormally large" trading volumes. 
Lillo et al. (2015) \cite{lillo2015news} analyzed the trading behavior of a large set of single investor of the Nokia stock and the role of endogenous factors (returns and volatility) and the exogenous factors from news. It has been shown that different categories of investors are differently correlated to these factors.
Vlastakis et al. (2016) \cite{vlastakis2012information} studied the information demand and supply with Google Trends database at the firm and market level using data for 30 of the largest stocks traded on NYSE and NASDAQ.

\end{document}